# GPU-parallelisation of wavelet-based grid adaptation for fast finite volume modelling: application to shallow water flows

Alovya Ahmed Chowdhury[1*], Georges Kesserwani[1], Charles Rougé[1], Paul Richmond[2]

[1]*Department of Civil and Structural Engineering, University of Sheffield, Mappin St, Sheffield, UK*

[2]*Department of Computer Science, University of Sheffield, Mappin St, Sheffield, UK*

E-mail: {aachowdhury2, g.kesserwani, c.rouge, p.richmond}@sheffield.ac.uk

*Corresponding author*

## Abstract

Wavelet-based grid adaptation driven by the "multiresolution analysis" (MRA) of the Haar wavelet (HW) allows to devise an adaptive first-order finite volume (FV1) model (HWFV1) that can readily preserve the modelling fidelity of its reference uniform-grid FV1 counterpart. However, the MRA incurs a high computational cost as it involves "encoding" (coarsening), "decoding" (refining), analysing and traversing modelled data across a deep hierarchy of nested, uniform grids. GPU-parallelisation of the MRA is needed to reduce its computational cost, but its algorithmic structure (1) hinders coalesced memory access on the GPU, and (2) involves an inherently sequential tree traversal problem. This work redesigns the algorithmic structure of the MRA in order to parallelise it on the GPU, addressing (1) by applying Z-order space-filling curves and addressing (2) by adopting a parallel tree traversal algorithm. This results in a GPU-parallelised HWFV1 model (GPU-HWFV1). GPU-HWFV1 is verified against its CPU predecessor (CPU-HWFV1) and its GPU-parallelised reference uniform-grid counterpart (GPU-FV1) over five shallow water flow test cases. GPU-HWFV1 preserves the modelling fidelity of GPU-FV1 while being up to 30 times faster. Compared to CPU-HWFV1, it is up to 200 times faster, suggesting the GPU-parallelised MRA could be used to speed up other FV1 models.

## 1. Introduction

Finite volume shallow water models numerically solve the two-dimensional shallow water equations, most of which are now parallelised on graphics processing units (GPUs) to deal with the computational cost of supporting real-scale flood modelling applications (Buttinger-Kreuzhuber et al., 2022; Carlotto et al., 2021; Caviedes-Voullième et al., 2023; Dazzi et al., 2020; Delmas & Soulaïmani, 2022; Ferrari et al., 2023; Gordillo et al., 2020; Han et al., 2022; Sanz-Ramos et al., 2023; Shaw et al., 2021; Xia et al., 2019). GPU-parallelised finite volume models using uniform grids have become the de facto standard in industrial hydraulic modelling packages (Flood Modeller 2D, 2022; InfoWorks ICM, 2018; MIKE 21 GPU, 2019), with one also using a non-uniform grid via static-in-time grid adaptation (TUFLOW HPC, 2018). However, the development of GPU-parallelised shallow water models with *dynamic*-in-time grid adaptation has received relatively less attention.

Classical adaptive mesh refinement (AMR; Berger & Colella, 1989) enforces grid adaptation by generating a non-uniform grid that uses finer cells only where higher resolution is needed, reducing the overall number of cells in the grid and thus the computational cost (Ghazizadeh et al., 2020; Gong et al., 2020; Holzbecher, 2022; Hu et al., 2018; Lakhlifi et al., 2018; Wallwork et al., 2020). A desirable goal in shallow water modelling is then to combine AMR and GPU-parallelisation in an effort to further reduce computational costs. However, the GPU-parallelisation of AMR is challenging due to its algorithmic complexity (de la Asunción & Castro, 2017; Qin et al., 2019): because of this, often, only the finite volume shallow water model is parallelised on the GPU, while the AMR operations for generating the non-uniform grid are retained on the central processing unit (CPU). Such a hybrid CPU-GPU approach, in turn, introduces computational overhead because data must be transferred between the CPU and the GPU. Other models eliminate this data transfer overhead by parallelising the AMR operations on the GPU as well; examples include applications in hydrodynamics (Beckingsale et al., 2015; Dunning et al., 2020; Sætra et al., 2014; Wahib et al., 2016), magnetohydrodynamics (Schive et al., 2018), combustion (Lung et al., 2016) and gas dynamics (Giuliani & Krivodonova, 2019). Nonetheless, the combination of AMR and GPU-parallelisation has not been as popular in shallow water modelling compared to the de facto standard GPU-parallelised uniform-grid shallow water models, as its design cannot guarantee the preservation of the modelling fidelity delivered by the reference uniform-grid finite volume shallow water model without AMR (Donat et al., 2014; Ghazizadeh et al., 2020; Kévin et al., 2017; Liang et al., 2015; Zhang et al., 2021; Zhou et al., 2013).

In contrast, wavelet-based grid adaptation is fundamentally designed on the basis of preserving a similar level of modelling fidelity initially deliverable by the reference uniform-grid counterpart, as demonstrated for shallow water models (Caviedes-Voullième & Kesserwani, 2015; Haleem et al., 2015). The design of the wavelet-based grid adaptation process requires specifying a single user-specified error threshold $\varepsilon$ that rigorously controls the deviation allowed to occur from the reference uniform-grid counterpart (Gerhard et al., 2015). However, wavelet-based grid adaptation imposes a much higher computational overhead than classical AMR as it uses the "multiresolution analysis" (MRA) to generate the non-uniform grid, which involves "encoding" (coarsening), "decoding" (refining), analysing and traversing modelled data across a deep hierarchy of nested, uniform grids. Hampered by the computational overhead of the MRA, shallow water models incorporating wavelet-based grid adaptation remain too costly to run for real-scale applications on the CPU (Kesserwani & Sharifian, 2020). Hence, the GPU-parallelisation of the MRA to reduce its computational overhead seems a necessary step forward to investigate if wavelet-based grid adaptation can accelerate the de facto standard GPU-parallelised finite volume shallow water models whilst preserving a similar level of modelling fidelity.

To date, wavelet-based grid adaptation has been mainly been parallelised on CPUs (Deiterding et al., 2020; Domingues et al., 2019; Gillis & van Rees, 2022; Julius & Marie, 2021; Semakin & Rastigejev, 2020; Soni et al., 2019; Zeidan et al., 2022). Its parallelisation on the GPU is unreported and remains necessary, as mentioned, to reduce the computational overhead of performing the MRA. In practice, this is difficult to achieve due to two obstacles. First, the modelled data involved in the MRA process are close together in the hierarchy of grids in physical space, but they are not guaranteed to be close together in GPU memory space unless the hierarchy is indexed in a deliberate way, hindering coalesced memory access which should be maximised to achieve fast GPU performance (Brodtkorb et al., 2013; NVIDIA, 2023). Second, traversing the hierarchy of grids boils down to a tree traversal problem, which is fundamentally a sequential task usually completed with recursive algorithms such as depth-first traversal (DFT; Sedgewick & Wayne, 2011).

This paper aims to overcome these two obstacles and implement wavelet-based grid adaptation on the GPU by combining two computational ingredients. This entails (1) introducing so-called space-filling curves (SFC; Bader, 2013; Sagan, 1994) to index the hierarchy of grids and ensure coalesced memory access, and (2) adopting a parallel tree traversal (PTT) algorithm (Karras, 2012) to replace the recursive DFT algorithm. To be of practical relevance, the GPU-parallelised wavelet-based grid adaptation process must make an adaptive wavelet-based shallow water model competitive with its GPU-parallelised reference uniform-grid counterpart.

The rest of this paper is organised as follows. Section 2 presents a GPU-parallelised adaptive Haar wavelet (HW) first-order finite volume (FV) shallow water model (Haleem et al., 2015; Kesserwani & Sharifian, 2020), termed GPU-HWFV1. Section 3 compares GPU-HWFV1 against the GPU-parallelised uniform-grid FV1 solver of the open-source LISFLOOD-FP 8.0 flood modelling package (Shaw et al., 2021), referred to hereafter as GPU-FV1. For completeness, GPU-HWFV1 is also compared against its sequential CPU predecessor, hereafter called CPU-HWFV1, which has already been extensively validated for one-dimensional and two-dimensional shallow water test cases (Kesserwani et al., 2019; Kesserwani & Sharifian, 2020). Section 4 presents conclusions on the potential benefits of using GPU-HWFV1 for shallow water modelling applications.

## 2. New GPU-HWFV1 model

This section presents a GPU-parallelised adaptive Haar wavelet (HW) first-order finite volume (FV1) shallow water model, GPU-HWFV1. GPU-HWFV1 is obtained by parallelising the sequential CPU version of the HWFV1 model presented in Kesserwani and Sharifian (2020), CPU-HWFV1. CPU-HWFV1 is difficult to parallelise because it incorporates a wavelet-based grid adaptation process whose algorithmic structure (1) hinders coalesced memory access on the GPU, and, (2) features an inherently sequential tree traversal problem. To clarify why this is so, Section 2.1 gives an overview of CPU-HWFV1 with a focus on presenting the grid adaptation process. Section 2.2 then presents the GPU-parallelisation of CPU-HWFV1, detailing in particular how two computational ingredients, namely, Z-order space-filling curves and a parallel tree traversal algorithm, are used to redesign the algorithmic structure of the grid adaptation process and make it suitable for GPU-parallelisation.

## 2.1 Existing sequential CPU-HWFV1 model

### 2.1.1 Overview

CPU-HWFV1 is an adaptive shallow water model that runs shallow water simulations by solving the two-dimensional shallow water equations (SWE) over a dynamically adaptive non-uniform grid. CPU-HWFV1 is made up of two mechanisms: a wavelet-based grid adaptation process for generating the non-uniform grid, and an FV1 scheme for updating the modelled data on this non-uniform grid. Since CPU-HWFV1 is dynamically adaptive, it performs the grid adaptation process every timestep before performing the FV1 update. In contrast to classical adaptive mesh refinement methods (e.g., Berger & Colella, 1989) which start with a coarse reference grid and selectively refine it, wavelet-based grid adaptation starts with a reference uniform grid at the finest allowable resolution and selectively coarsens it. The wavelet-based grid adaptation process is driven by the "multiresolution analysis" (MRA) of the HW (Haleem et al., 2015; Kesserwani & Sharifian, 2020) which starts on a reference uniform grid made up of $2^L \times 2^L$ cells, where $L$ is a user-specified integer denoting the maximum refinement level that controls the resolution of the finest grid.

The conservative form of the SWE are initially discretised over this reference uniform grid, which can be written as:

$$\partial_t \mathbf{U} + \partial_x \mathbf{F}(\mathbf{U}) + \partial_y \mathbf{G}(\mathbf{U}) = \mathbf{S}_b(\mathbf{U}) + \mathbf{S}_f(\mathbf{U}). \tag{1}$$

Where $\partial_t$, $\partial_x$ and $\partial_y$ represent partial derivatives with respect to $t$, $x$ and $y$, respectively. The vector $\mathbf{U} = [h, hu, hv]^T$ contains the flow variables, $\mathbf{F} = [hu, (hu)^2/h + 1/2gh^2, huv]^T$, $\mathbf{G} = [hv, huv, (hv)^2/h + 1/2gh^2]^T$ are the components of the flux vector, and $\mathbf{S}_b = [0, -gh\partial_x z, -gh\partial_y z]^T$ and $\mathbf{S}_f = [0, -C_f u\sqrt{u^2+v^2}, -C_f v\sqrt{u^2+v^2}]^T$ are source term vectors. The variable $h(x, y, t)$ is the water depth (m), $u(x, y, t)$ and $v(x, y, t)$ are the $x$- and $y$-components of the velocity (m/s), respectively, and $g$ is the gravitational acceleration constant (m/s$^2$). In $\mathbf{S}_b$, $z(x, y)$ is the topographic elevation (m), and in $\mathbf{S}_f$, $C_f = gn_M^2/h^{1/3}$ where $n_M$ is Manning's coefficient (s$^{-1}$m$^{1/3}$).

The FV1 scheme locally approximates the variables $h$, $hu$, $hv$ and $z$ as piecewise-constant modelled data over each cell in the grid, for which scalar coefficients $h_c$, $(hu)_c$, $(hv)_c$ and $z_c$ are assigned per cell. Ordinarily, on a uniform grid, an FV1 update is performed at every timestep using a forward-Euler scheme applied to a spatial operator $\mathbf{L}_c$ to advance the vector of flow coefficients $U_c = [h_c, (hu)_c, (hv)_c]^T$ in time as follows:

$$\partial_t \mathbf{U}_c(t) = \mathbf{L}_c. \tag{2}$$

To compute $\mathbf{L}_c$, the local coefficients $S_c = \{h_c, (hu)_c, (hv)_c, z_c\}$ as well as the coefficients of the neighbour cells, denoted by $S_{west}$, $S_{east}$, $S_{north}$, and $S_{south}$, are needed. The expression for $\mathbf{L}_c$ is established in previous studies (Liang, 2010) and is therefore outside of the scope of this paper. In summary, $\mathbf{L}_c$ involves flux calculations using a Harten-Lax-van Leer Riemann solver, and also includes treatments to ensure the positivity of the water depth and the so-called well-balancedness of the FV1 scheme over wet/dry zones and fronts.

### 2.1.2 Wavelet-based grid adaptation process driven by MRA

At every timestep, before performing the FV1 update, CPU-HWFV1 performs an MRA to generate a non-uniform grid that is adapted to the details of the flow and topography. The MRA involves a hierarchy of nested grids of increasingly coarser resolution relative to the reference $2^L \times 2^L$ grid at the maximum refinement level $L$. Figure 1a shows a hierarchy of grids with $L = 2$. The refinement level of each grid in the hierarchy is denoted by $n$: the finest grid in the hierarchy is at refinement level $n = L = 2$, the second-finest grid is at refinement level $n = 1$ and the coarsest grid is at $n = 0$. Let $s^{(n)}$ denote the coefficients $s$ at refinement level $n$, where $s$ denotes any of the quantities in the set $S_c$ for ease of presentation of the MRA.

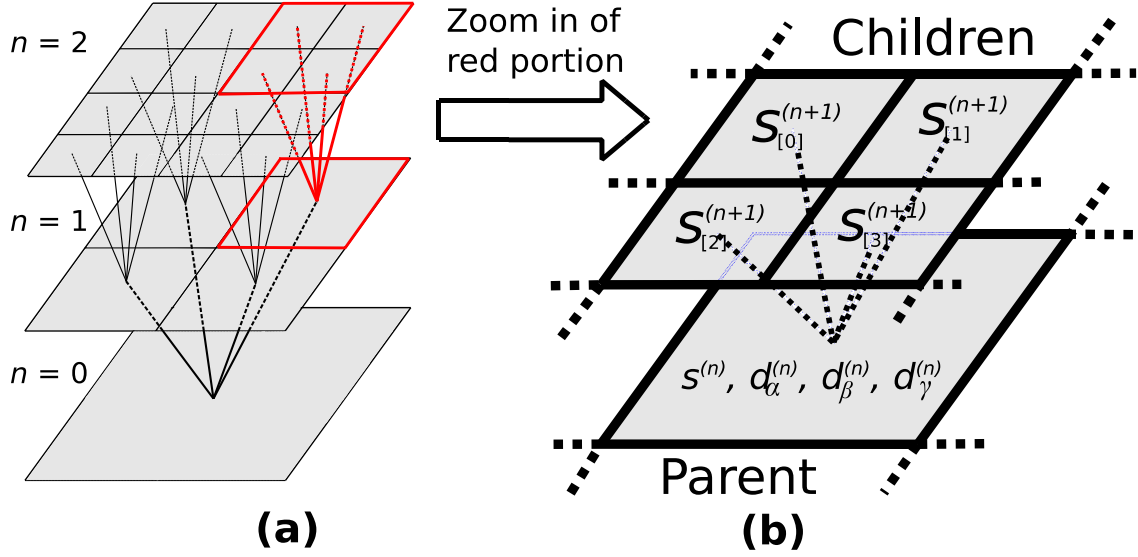

Figure 1: Multiresolution analysis (MRA). Left panel shows a hierarchy of grids involved in the MRA, with a maximum refinement level $L = 2$. Right panel shows how four cells at refinement level $n + 1$, called "children", are related to a single cell at refinement level $n$, called "parent". Also shown are the coefficients $s$ and details $d$ that are involved in computations to realise the MRA.

**Encoding**. At the start of the MRA process, only $s^{(L)}$ are available, corresponding to the coefficients obtained from initially discretising the SWE over the reference uniform grid. The MRA process continues by producing $s^{(n)}$ at the lower refinement levels, i.e., for $n = L - 1, L - 2, \ldots, 1, 0$. The coefficient $s^{(n)}$ of a cell at refinement level $n$ (called "parent") is produced using the coefficients $s^{(n+1)}_{[0]}$, $s^{(n+1)}_{[1]}$, $s^{(n+1)}_{[2]}$ and $s^{(n+1)}_{[3]}$ of four cells at refinement level $n + 1$ (called "children"), in particular by applying Eq. 3a[1]. Figure 1b shows the parent-children stencil dictating how the children $s^{(n+1)}_{[0]}$, $s^{(n+1)}_{[1]}$, $s^{(n+1)}_{[2]}$ and $s^{(n+1)}_{[3]}$ at level $n + 1$ are positioned relative to the parent $s^{(n)}$ at level $n$. Equations 3b - 3d are also applied to produce so-called "details", denoted by $d^{(n)}_\alpha$, $d^{(n)}_\beta$ and $d^{(n)}_\gamma$. These details are used to decide which cells to include in the non-uniform grid based on a normalised detail $d^{(n)}_{norm} = max(d^{(n)}_\alpha, d^{(n)}_\beta, d^{(n)}_\gamma)/s_{max}$, where $s_{max}$ is the largest coefficient for all $s^{(L)}$. Cells whose $d^{(n)}_{norm}$ is greater than $2^{n-L}\varepsilon$ are deemed to have significant details, where $\varepsilon$ is a user-specified error threshold.

---

[1] In Eqs. 3a - 4d, $H^0$, $H^1$, $G^0$ and $G^1$ are scalar coefficients obtained from the Haar wavelets whose derivation is available in previous works (    Kesserwani & Sharifian, 2020; Keinert, 2003), and is outside of the scope of this paper.

$$s^{(n)} = H^0(H^0 s_{[0]}^{(n+1)} + H^1 s_{[2]}^{(n+1)}) + H^1(H^0 s_{[1]}^{(n+1)} + H^1 s_{[3]}^{(n+1)}) \quad (3a)$$

$$d_\alpha^{(n)} = H^0(G^0 s_{[0]}^{(n+1)} + G^1 s_{[2]}^{(n+1)}) + H^1(G^0 s_{[1]}^{(n+1)} + G^1 s_{[3]}^{(n+1)}) \quad (3b)$$

$$d_\beta^{(n)} = G^0(H^0 s_{[0]}^{(n+1)} + H^1 s_{[2]}^{(n+1)}) + G^1(H^0 s_{[1]}^{(n+1)} + H^1 s_{[3]}^{(n+1)}) \quad (3c)$$

$$d_\gamma^{(n)} = G^0(G^0 s_{[0]}^{(n+1)} + G^1 s_{[2]}^{(n+1)}) + G^1(G^0 s_{[1]}^{(n+1)} + G^1 s_{[3]}^{(n+1)}) \quad (3d)$$

The process of computing $s^{(n)}$, $d_\alpha^{(n)}$, $d_\beta^{(n)}$ and $d_\gamma^{(n)}$ at the lower refinement levels is called "encoding". Algorithm 1 shows pseudocode summarising the process of encoding. The pseudocode has an outer loop (lines 2 to 10) and an inner loop (lines 3 to 9). The outer loop iterates over each grid in the hierarchy, starting from the grid at the second-highest refinement level, $n = L - 1$, to the grid at the lowest refinement level, $n = 0$. The inner loop iterates through each cell in the grid while applying Eqs. 3a - 3d (lines 5 and 6) and flagging significant details (line 8).

```
1  algorithm ENCODING(L, ε)
2    for each grid in hierarchy from L - 1 to 0 do
3      for each cell in the grid do
4        load children s_[0]^(n+1), s_[1]^(n+1), s_[2]^(n+1), s_[3]^(n+1)   for Eqs. 3a - 3d
5        compute parent coefficient s^(n) using Eq. 3a
6        compute details d_α, d_β, d_γ using Eqs. 3b -    3d
7        compute normalised detail d_norm
8        flag as significant if d_norm ≥ 2^(n-L)ε
9      end for
10     end for
11 end algorithm
```

Algorithm 1: Pseudocode for the process of computing $s^{(n)}$, $d_\alpha^{(n)}$, $d_\beta^{(n)}$ and $d_\gamma^{(n)}$ at the lower refinement levels, called "encoding".

**Decoding**. Flagging significant details during the encoding process results in a tree-like structure of significant details that CPU-HWFV1 uses to identify the cells that make up the non-uniform grid. Figure 2a shows an example of a tree of significant details. CPU-HWFV1 traverses the tree starting from the coarsest cell while applying Eqs. 4a - 4d and stops climbing whenever it reaches either a cell on the finest grid or a cell with a detail that is not significant. This way of traversing the tree corresponds to the application of a depth-first traversal (DFT) algorithm (Sedgewick & Wayne, 2011). The path of the DFT can be traced in Figure 1a by following the cells labelled from 0 to 12 in ascending order. The cells that CPU-HWFV1 visits during the DFT where the tree terminates (i.e., when CPU-HWFV1 visits a cell on the finest grid or a cell with a detail that is not significant) are identified as "leaf" cells (indicated in blue and green in Figure 1a). CPU-HWFV1 assembles the identified leaf cells into a non-uniform grid. Figure 1b shows the non-uniform grid generated by CPU-HWFV1 after assembling the leaf cells in Figure 1a.

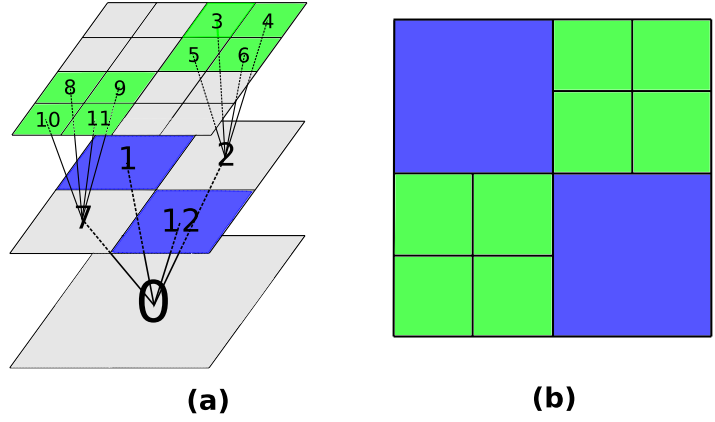

Figure 2: Left panel shows the tree-like structure obtained after flagging significant details during the process of encoding; the cells where the tree terminates are called "leaf" cells (highlighted in green and blue). Right panel shows how the leaf cells are assembled into a non-uniform grid.

$$s_{[0]}^{(n+1)} = H^0(H^0 s^{(n)} + G^0 d_\alpha^{(n)}) + G^0(H^0 d_\beta^{(n)} + G^0 d_\gamma^{(n)}) \tag{4a}$$

$$s_{[2]}^{(n+1)} = H^0(H^1 s^{(n)} + G^1 d_\alpha^{(n)}) + G^0(H^1 d_\beta^{(n)} + G^1 d_\gamma^{(n)}) \tag{4b}$$

$$s_{[1]}^{(n+1)} = H^1(H^0 s^{(n)} + G^0 d_\alpha^{(n)}) + G^1(H^0 d_\beta^{(n)} + G^0 d_\gamma^{(n)}) \tag{4c}$$

$$s_{[3]}^{(n+1)} = H^1(H^1 s^{(n)} + G^1 d_\alpha^{(n)}) + G^1(H^1 d_\beta^{(n)} + G^1 d_\gamma^{(n)}) \tag{4d}$$

The process of performing a DFT while applying Eqs. 4a - 4d and identifying leaf cells is called "decoding". Algorithm 2 shows pseudocode describing the decoding process. The algorithm launches at the coarsest cell in the tree. The children $s_{[0]}^{(n+1)}, s_{[1]}^{(n+1)}, s_{[2]}^{(n+1)}$ and $s_{[3]}^{(n+1)}$ of this cell are computed using Eqs. 4a - 4d (line 6) and then the algorithm is relaunched using the children coefficients at one refinement level higher (lines 7 to 10). The algorithm is recursively launched unless a cell with a detail that is not significant is reached or a cell on the finest grid is reached (line 2), at which point the cell is identified as a leaf cell (line 3).

1  **recursive algorithm** DECODING($s^{(n)}$, $n$)
2   **if** detail is not significant **or** reached finest grid **then**
3     identify cell as leaf cell
4     stop decoding
5   **else**
6     compute children $s_{[0]}^{(n+1)}, s_{[1]}^{(n+1)}, s_{[2]}^{(n+1)}, s_{[3]}^{(n+1)}$  with Eqs. 4a - 4d
7     DECODING($s_{[0]}^{(n+1)}$, $n + 1$)
8     DECODING($s_{[1]}^{(n+1)}$, $n + 1$)
9     DECODING($s_{[2]}^{(n+1)}$, $n + 1$)
10    DECODING($s_{[3]}^{(n+1)}$, $n + 1$)
11   **end if**
12 **end recursive algorithm**

Algorithm 2: Pseudocode for performing a depth-first traversal of the tree of significant details (obtained after encoding) while applying Eqs. 4a - 4d, called "decoding".

### 2.1.3 Neighbour finding to perform the FV1 update over the non-uniform grid

Decoding allows CPU-HWFV1 to identify leaf cells and assemble them into a non-uniform grid, over which the FV1 update is performed. To compute $\mathbf{L}_c$ for performing the FV1 update, CPU-HWFV1 needs to retrieve the sets of coefficients of the neighbours $S_{west}$, $S_{east}$, $S_{north}$ and $S_{south}$ for each leaf cell. Retrieving $S_{west}$, $S_{east}$, $S_{north}$ and $S_{south}$ is trivial on a uniform grid because a cell can look left, right, up and down to find its neighbours, but this is not as straightforward on a non-uniform grid, since a cell can have multiple neighbours in a given direction. Figure 3 shows an example of a cell and its neighbours in a non-uniform grid. In this example, finding $S_{west}$ (blue cell), $S_{south}$ and $S_{north}$ (grey cells) is straightforward. However, it is not clear what $S_{east}$ should be because the eastern neighbours (red cells) are at a higher refinement level and there are multiple neighbours. CPU-HWFV1 avoids this confusion by taking $S_{east}$ to be the set of coefficients of the eastern neighbour at the same refinement level (yellow cell in Figure 3), which is readily available since the encoding and decoding processes have produced $s^{(n)}$ at all refinement levels. Being able to find the neighbours for retrieving $S_{west}$, $S_{east}$, $S_{north}$ and $S_{south}$ makes it trivial for CPU-HWFV1 to compute $\mathbf{L}_c$ and perform the FV1 update per cell.

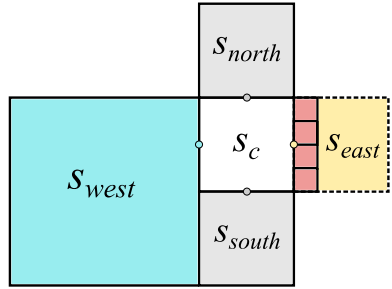

Figure 3: Finding the neighbours of a cell in a non-uniform grid to retrieve the sets $S_c$, $S_{west}$, $S_{east}$, $S_{north}$ and $S_{south}$ in order to compute the spatial operator $\mathbf{L}_c$.

This completes the description of the steps involved in the CPU-HWFV1 model, and Algorithm 3 shows pseudocode summarising CPU-HWFV1. To run CPU-HWFV1, the user needs to specify the maximum refinement level ($L$), the error threshold ($\varepsilon$) and the simulation end time ($t_{end}$) (line 1). CPU-HWFV1 runs in a loop until $t_{end}$ is reached (lines 3 to 10). A non-uniform grid is generated every iteration of the loop, i.e., at every timestep (lines 4 to 6). Note that after the first timestep, the details are zeroed first before re-encoding, and encoding is performed only along the tree of significant details. On this non-uniform grid, CPU-HWFV1 performs an FV1 update (line 7), after which the simulation time is incremented by the current timestep (line 8) while a new timestep is recomputed based on the Courant-Friedrich-Lewy (CFL) condition (line 9); for stability, a CFL number of 0.5 is used (Kesserwani & Sharifian, 2020).

```
1  algorithm HWFV1(L, ε, t_end)
2     get s^(L) from initial discretisation of SWE on finest grid
3     while current time < t_end do
4        ENCODING(L, ε)
5        DECODING(s^(0), 0) // start decoding from coarsest cell
6        find neighbours to compute L_c
7        use L_c to perform FV1 update
8        increment current time by timestep
9        compute new timestep based on CFL condition
10    end while
11 end algorithm
```

Algorithm 3: Pseudocode summarising the steps involved in the HWFV1 model.

## 2.2 GPU-parallelisation of CPU-HWFV1

To obtain GPU-HWFV1, the CPU-HWFV1 model, summarised in Algorithm 3, is parallelised on an NVIDIA GPU using CUDA. In the CUDA programming model, instructions are executed in parallel by threads, and a group of 32 threads that operate in lockstep is known as a warp. When parallelising on a GPU using CUDA, two important considerations are to maximise coalesced memory access and minimise warp divergence (Brodtkorb et al., 2013, NVIDIA, 2023). Coalesced memory access occurs when adjacent threads within a warp access contiguous memory locations. Warp divergence occurs threads within a warp      execute different instructions. A naive parallelisation of the steps in Algorithm 3 would not properly address the issues of coalesced memory access and/or warp divergence. Consider the parallelisation of the encoding process (line 4 of Algorithm 3; Algorithm 1), assuming the cells in the hierarchy of grids are naively indexed in row-major order. Figure 4 shows the hierarchy of grids in Figure 1a indexed in row-major order (left panel) and the corresponding locations of the children $s_{[0]}^{(n+1)}, s_{[1]}^{(n+1)}, s_{[2]}^{(n+1)}$ and $s_{[3]}^{(n+1)}$ in memory (right panel). Using row-major indexing leads to uncoalesced memory access when loading the children (line 3 of Algorithm 1) because there are gaps between the memory locations.

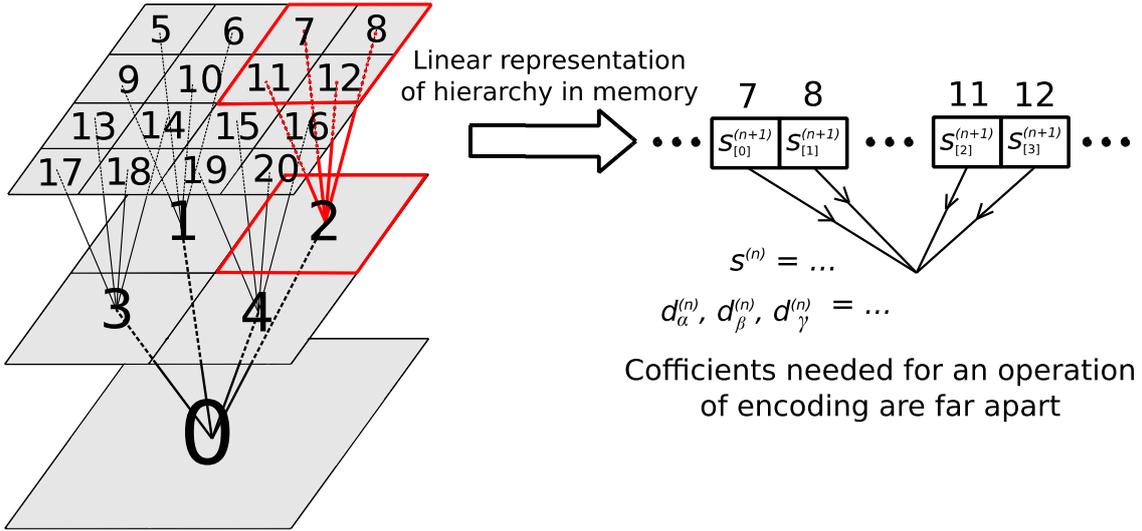

Figure 4: Indexing the hierarchy of grids in row-major order (left panel) and the corresponding locations of the children $s_{[0]}^{(n+1)}, s_{[1]}^{(n+1)}, s_{[2]}^{(n+1)}$ and $s_{[3]}^{(n+1)}$ in memory (right panel).

### 2.2.1 Ensuring coalesced memory access in encoding and decoding via Z-order curves

To facilitate coalesced memory access, a different type of indexing is needed that ensures that coefficients that are nearby in the grid are also nearby in memory. Space-filling curves (SFCs) allow this kind of indexing by definition because they map spatial data to a one-dimensional line such that data close together in space tend to be close together on the line (Bader, 2013; Sagan, 1994). There are a few different types of SFCs such as the Sierpinski curve, the Peano curve, the Hilbert curve and the Z-order curve (also known as Lebesgue or Morton curve). All of these SFCs have been previously used in the context of adaptive mesh refinement (Brix et al., 2009; Burstedde et al., 2011; Meister et al., 2016; Weinzierl & Mehl, 2011), and also once in the context of wavelet-based grid adaptation (Brix et al., 2009), but none of these works involved GPU-parallelisation. To the authors' knowledge, this is the first work to use a SFC to implement wavelet-based grid adaptation on the GPU. In particular, the Z-order SFC is chosen because its motif matches exactly with the parent-children square stencil shown in Figure 1b.

A Z-order curve can be created for a square $2^n \times 2^n$ grid by following the so-called Morton codes of each cell in the grid in order. The Morton code of a cell is obtained by interleaving the bits of its $i$ and $j$ positional indices in the grid. Figure 5 shows the creation of a Z-order curve for a $2^2 \times 2^2$ grid: the left panel shows the $i$ (black) and $j$ (red) indices of each cell in binary form and how the bits are interleaved (alternating red and black) to yield Morton codes (in binary form). The right panel shows how these Morton codes (in decimal form) are followed in ascending order to create the Z-order curve. The curve allows to enforce Z-order indexing of the grid because each cell in the grid can be identified on the curve via its (unique) Morton code.

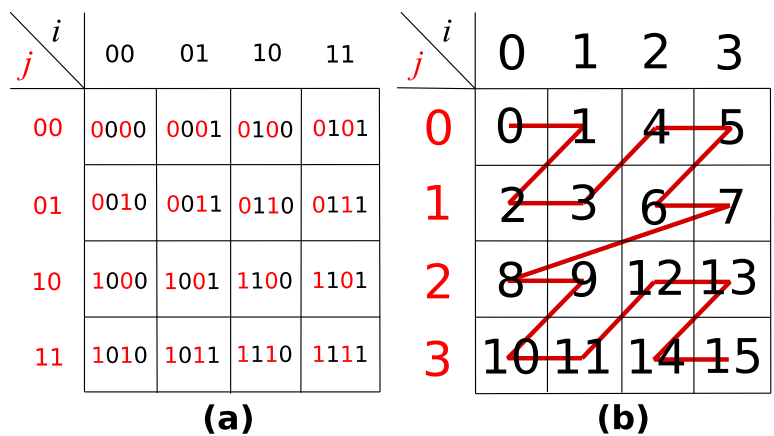

Figure 5: Creation of a Z-order curve for a $2^2 \times 2^2$ grid. The left panel shows how the binary forms of the $i$ and $j$ indices of each cell making up a $2^2 \times 2^2$ grid are bit interleaved (alternating red and black digits) to yield so-called Morton codes (also in binary). The right panel shows how these Morton codes (in decimal form) are followed in ascending order to create a Z-order curve and enforce Z-order indexing of the grid.

In GPU-HWFV1, Z-order curves are created for each grid in the hierarchy while enforcing continuity in the indexing of the curves of subsequent grids, resulting in a unique Z-order index for each cell in the hierarchy. Figure 6 shows Z-order indexing of the cells in the hierarchy of grids in Figure 1a, alongside the corresponding locations of the children $s_{[0]}^{(n+1)}, s_{[1]}^{(n+1)}, s_{[2]}^{(n+1)}$ and $s_{[3]}^{(n+1)}$ in memory. Enforcing this Z-order indexing allows for coalesced memory access during encoding because the children are in contiguous memory locations, as seen in the right panel of Figure 6.

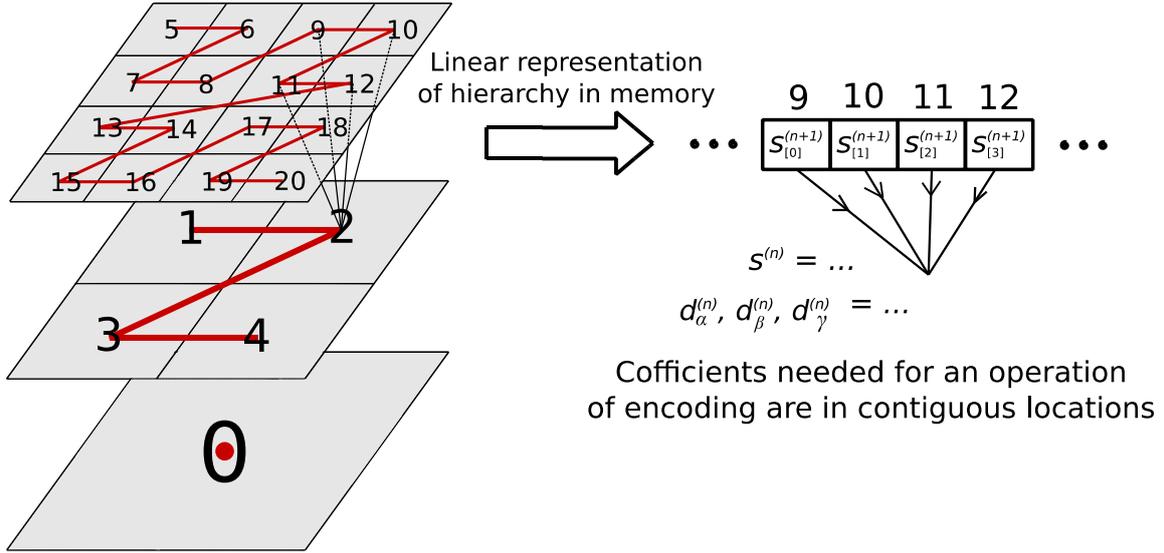

Figure 6: Z-order indexing of the hierarchy of grids so that each cell in the hierarchy has a unique Z-order index (left panel) and the corresponding locations of the children in memory (right panel).

Parallelising the decoding process (line 5 of Algorithm 3; Algorithm 2) is more difficult than parallelising the encoding process because decoding involves the inherently sequential DFT algorithm. To overcome this difficulty, decoding is broken down into two parts that are parallelised separately: the first part is the application of Eqs. 4a - 4d and the second part is the identification of leaf cells. *Hereafter, decoding refers exclusively to the application of Eqs. 4a - 4d, not the identification of leaf cells.*

Decoding can be parallelised relatively easily because it can be performed using loops (similar to those in Algorithm 1) instead of using DFT. Algorithm 4 shows pseudocode describing how to perform decoding in parallel. The pseudocode involves an outer loop and a parallelised inner loop. The outer loop iterates over the grids in the hierarchy starting from the coarsest grid up to the second-finest grid (lines 2 to 9) while the inner loop iterates through each cell in the grid in parallel (lines 3 to 8). The inner loop checks if the detail is significant (line 4), loads the parent $s^{(n)}$ and the details $d_\alpha^{(n)}$, $d_\beta^{(n)}$ and $d_\gamma^{(n)}$ (line 5) and computes the children $s_{[0]}^{(n+1)}, s_{[1]}^{(n+1)}, s_{[2]}^{(n+1)}$ , $s_{[3]}^{(n+1)}$ (line 6). This parallel inner loop, in particular the loading of the children coefficients in line 5, has coalesced memory access because of Z-order indexing; this can be seen by interpreting the arrows in Figure 6 in the reverse direction.

```
1  algorithm PARALLEL_DECODING(L)
2    for each grid in hierarchy from 0 to L - 1 do
3      for each cell in the grid do in parallel
4        if detail is significant then
5          load parent s^(n) and details d_α^(n), d_β^(n), d_γ^(n)
6          compute children s_[0]^(n+1), s_[1]^(n+1), s_[2]^(n+1), s_[3]^(n+1)   with Eqs. 4a - 4d
7        end if
8      end for
9    end for
10 end algorithm
```

Algorithm 4: Pseudocode for the decoding process in parallel.

### 2.2.2 Parallel tree traversal algorithm

Unlike decoding, the identification of leaf cells boils down to a tree traversal problem. There have been many investigations into parallel tree traversal (PTT) algorithms on the GPU (Bédorf et al., 2012; Chitalu et al., 2018; Goldfarb et al., 2013; Karras, 2012; Lohr, 2009; Nam et al., 2016; Zola et al., 2014), hinting that the identification of leaf cells can be parallelised by adopting a PTT algorithm. In this work, a modified version of the PTT algorithm developed by Karras (2012) is adopted because it can be easily modified to work with Z-order indexing. The PTT algorithm traverses the tree of significant details as follows, explained by example considering the tree in Figure 2a without loss of generality.

Figure 7a shows the tree after enforcing Z-order indexing of the hierarchy of grids, and different traversal paths are highlighted in cyan, magenta, yellow and grey. The PTT algorithm starts by launching as many threads as there are cells on the finest grid ($2^2 \times 2^2 = 16$). Each thread tries to reach a target cell on the finest grid by traversing progressively finer cells. A thread is denoted by $t_m$, where $m$ is the thread index (here $m = 0, 1, \ldots, 15$). The target cell of $t_m$ is the cell on the finest grid with a Morton code with the same thread index $m$, e.g., $t_3$ has thread index 3 and tries to reach the cell on the finest grid with Morton code 3 [2]. A thread stops if it either reaches this target cell or encounters a cell with a detail that is not significant (analogous to identifying a leaf cell). The thread then records the Z-order index of the cell at which it stops. Figure 7b shows the traversal paths of each thread during PTT in terms of the Z-order indices of the cells they traverse. These paths show that divergence is greatly minimised because adjacent threads perform similar traversals. For example, $t_0$ to $t_3$ are adjacent threads and they follow the same cyan path in Figure 7a. Similarly, $t_4$ to $t_7$ follow the magenta path, $t_8$ to $t_{11}$ follow the yellow path and $t_{12}$ to $t_{15}$ follow the grey path. Figure 7c shows the Z-order indices recorded by each thread after PTT is complete. These Z-order indices correspond to the indices of leaf cells.

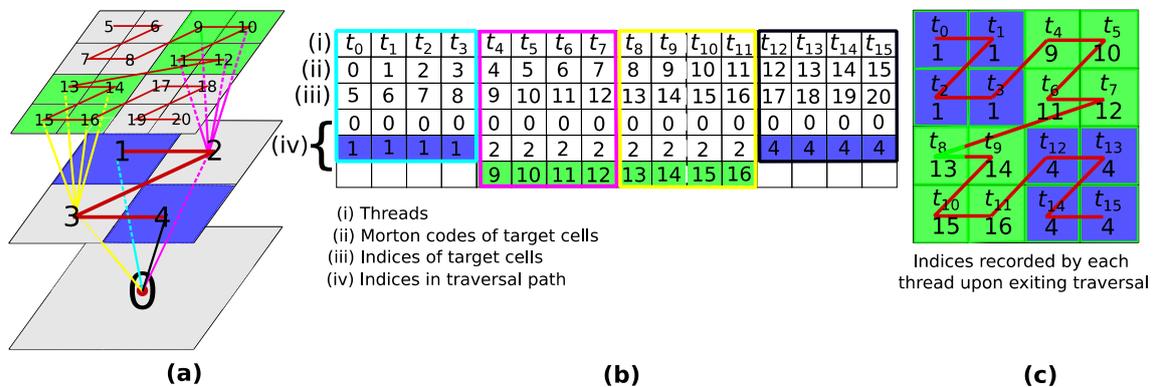

**(a)** **(b)** **(c)**

Figure 7: Parallel tree traversal (PTT). The left panel shows the tree of significant details after enforcing Z-order indexing, with different traversal paths indicated in yellow, cyan, magenta and grey. The middle panel shows the traversal paths of each thread during the PTT in terms of the Z-order indices of the cells they traverse. The right panel shows the Z-order indices recorded by each thread after PTT is complete.

---

[2] Note that the Morton code refers to the index of a cell in a single grid, whereas the Z-order index refers to the index of a cell within the hierarchy. For example, the cell on the finest grid with Morton code 3 (see Figure 5b) has a Z-order index in the hierarchy of 8 (see Figure 6).

```
1  algorithm PARALLEL_TREE_TRAVERSAL
2    for each cell in finest grid do in parallel
3      start at coarsest cell
4      while try to reach finer cell do
5        if detail is not significant then
6          record z-order index of cell
7          stop traversing
8        else
9          if reached finest cell then
10           record z-order index of cell
11           stop traversing
12         else
13           try to reach finer cell
14         end if
15       end if
16     end while
17   end for
19 end algorithm
```

Algorithm 5: Pseudocode for parallel tree traversal (PTT) of the tree of significant details. The PTT features an iterative procedure (lines 4 to 16) instead of the recursive procedure in the depth-first traversal (DFT) in Algorithm 2.

### 2.2.3 Neighbour finding and FV1 update

Some of the indices recorded after PTT are duplicates because some threads ($t_0$ to $t_3$ and $t_{12}$ to $t_{15}$) finish at the same leaf cell. These duplicates are used to record the Z-order indices of the neighbour cells (which is necessary for the FV1 update) in parallel (line 6 of Algorithm 3) by making each thread in the grid in Figure 7c look left, right, up and down. The Z-order indices of the leaf cells and their neighbours are stored in memory. Figure 8a shows how the Z-order indices of the leaf cells and their neighbours are stored in five contiguous arrays. Duplicate groups of indices are removed as indicated by the double black lines via so-called parallel stream compaction using the CUB library (Merrill, 2022). Figure 8b shows the Z-order indices of the leaf cells and their neighbours without any duplicates, stored in five arrays. The leaf cells make up a non-uniform grid.

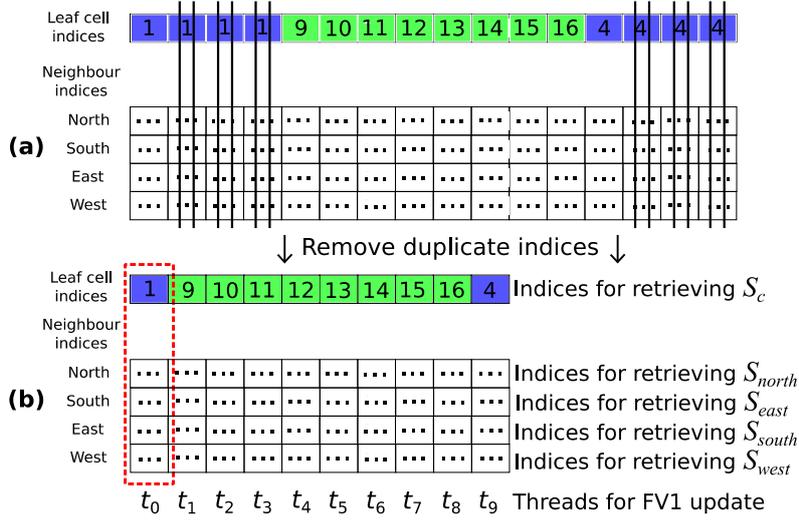

Figure 8: Parallel FV1 update. Top panel shows the Z-order indices of the leaf cells and their neighbours stored in memory after PTT and neighbour finding. Bottom panel shows the Z-order indices of the leaf cells and their neighbours without any duplicates, used to retrieve $S_c$, $S_{west}$, $S_{east}$, $S_{north}$ and $S_{south}$ to compute $\mathbf{L}_c$ and perform the FV1 update.

The next step is to parallelise the FV1 update on the leaf cells making up the non-uniform grid (line 7 of Algorithm 3) which is relatively simple. The parallel FV1 update launches one thread per leaf cell. In Figure 8b, there are ten leaf cells, so ten threads $t_0$ to $t_9$ are launched. Each thread uses the arrays of leaf cell and neighbour Z-order indices to retrieve $S_c$, $S_{west}$, $S_{east}$, $S_{north}$ and $S_{south}$ from within the hierarchy. For example, $t_0$ would use the first column of indices (red box in Figure 8b), $t_1$ would use the next column, and so on. Since each thread $t_0$ to $t_9$ retrieves $S_c$, $S_{west}$, $S_{east}$, $S_{north}$ and $S_{south}$ for each leaf cell, $\mathbf{L}_c$ can be computed to perform the FV1 update in parallel for all leaf cells. After the FV1 update, incrementing the current simulation time by the timestep (line 8 of Algorithm 3) is trivial. The new minimum timestep based on the CFL condition (line 9 of Algorithm 3) is computed by performing a so-called parallel minimum reduction using the CUB library (Merrill, 2022). Hence, the steps involved in the CPU-HWFV1 model (lines 4 to 9 of Algorithm 3) are all parallelised and GPU-HWFV1 is obtained.

## 3. Numerical results and discussion

This section will compare the proposed GPU-HWFV1 model against two existing and validated shallow water models, namely, the reference uniform-grid FV1 counterpart parallelised on the GPU, GPU-FV1, available in the open-source LISFLOOD-FP 8.0 flood modelling package (Shaw et al., 2021), and the sequential CPU predecessor (Kesserwani and Sharifian, 2020), CPU-HWFV1. The comparisons will be performed for five test cases (see Table 1) that explore a range of topographies and flow dynamics. The primary aim of the comparisons will be to quantify GPU-HWFV1's runtime performance relative to GPU-FV1 and CPU-HWFV1. A secondary aim is to verify that the proposed GPU-HWFV1 model is valid, which is done by checking that GPU-HWFV1 preserves a similar level of fidelity as GPU-FV1 at the finest resolution accessible to GPU-HWFV1.

Table 1: List of the test cases used to benchmark GPU-HWFV1.

| Test name | Test type | Reason for use | Previously used in |
|---|---|---|---|
| Quiescent flow over irregular topographies with different steepness. Dam-break flow over realistic terrain with friction effects (Section 3.1) | Synthetic | Verifying fidelity | (Huang et al., 2013; Kesserwani et al., 2018; Kesserwani & Sharifian, 2020; Shirvani et al., 2021; Song et al., 2011) |
| Circular 2D dam-break flow (Section 3.2) | Synthetic | Assessing runtime performance | (Kesserwani & Sharifian, 2020; Wang et al., 2011) |
| Pseudo-2D dam-break flow (Section 3.2) | Synthetic | Assessing runtime performance | (Kesserwani et al., 2019; Kesserwani & Sharifian, 2020) |
| Dam-break wave interaction with an urban district (Section 3.3) | Experimental | Assessing runtime performance | (Jeong et al., 2012; Caviedes-Voullième et al., 2020; Jeong et al., 2012; Kesserwani & Sharifian, 2020) |
| Tsunami wave propagation over a | Experimental | Assessing runtime | (Arpaia & Ricchiuto, |

| complex beach (Section 3.3) | performance | 2018; Caviedes-Voullième et al., 2020; Hou et al., 2015; Kesserwani & Sharifian, 2020) |

The first test case (Section 3.1) will focus solely on verifying GPU-HWFV1's fidelity under idealised scenarios where a benchmark solution is available. The second and third test cases (Section 3.2) will then mainly focus on assessing GPU-HWFV1's runtime performance against CPU-HWFV1 and GPU-FV1 in more complex cases considering synthetic dam break flows over flat terrain. As there is no terrain data to consider in these scenarios, the fidelity of GPU-HWFV1 will be assessed by verifying that its results match those from CPU-HWFV1 and GPU-FV1. This then allows to systematically analyse the runtime performance in relation to the sensitivity to triggering grid refinement ($\varepsilon$), the depth in grid resolution ($L$), and the flow type (vigorous or smooth). Therefore, simulations are run by pairing different values for $\varepsilon$ and $L$, with $\varepsilon = \{10^{-4}, 10^{-3}, 10^{-2}\}$ to cover the recommended ranges for maintaining a fair balance between the predictive accuracy and runtime performance (Kesserwani et al., 2019; Kesserwani & Sharifian, 2020) and, $L = \{8, 9, 10, 11\}$ as no gains in runtime performance was identified for $L \leq 7$ and using $L \geq 12$ exceeded the memory capacity of the GPU card used (RTX 2070). The effects of these parameters and the flow type on the wavelet-based grid adaptation of the HWFV1 models is summarised in Table 2.

Table 2: Aspects against which the runtime performance of GPU-HWFV1 over CPU-HWFV1 and over GPU-FV1 are assessed.

| Aspects | Description | Finest grid resolution |
|---------|-------------|------------------------|
| $L$ | Controls the finest accessible grid resolution | Deeper with higher $L$ |
| $\varepsilon$ | Controls how far the finest grid resolution is accessed | More accessible with smaller $\varepsilon$ |
| Flow | Vigorous (with discontinuities) to smooth (up to flat) | Triggered often for vigorous flows |

The expectations on runtime performance established from the synthetic test cases will finally be explored in the fourth and fifth test cases (Section 3.3), which involve realistic topographies represented by Digital Elevation Models (DEM). Running the HWFV1 models with the presence of a DEM means that the maximum refinement $L$ must be set to accommodate the DEM resolution, and no grid coarsening is allowed beyond what the MRA of the DEM suggests.

## 3.1 Verification of fidelity

The first synthetic test case in Table 1 is considered to verify the fidelity of GPU-HWFV1 under two simulation scenarios. The first scenario is to verify GPU-HWFV1's ability to preserve a quiescent (i.e., unmoving) flow state in the presence of wet-dry fronts with different levels of steepness in topography and different wetting conditions. This is done by checking its well-balancedness, which is the ability to balance the fluxes with the source terms under steady-state conditions (Greenberg & Leroux, 1996). If GPU-HWFV1 is well-balanced, then it should not spuriously disturb an initially quiescent flow state as the fluxes should exactly balance with the source terms. The second simulation scenario is to verify GPU-HWFV1's ability to reproduce a realistic dam-break flow with friction effects and moving wet-dry fronts. For both scenarios, the domain area is 70 m × 30 m with closed wall boundaries and includes humps to represent an irregular topography profile. To verify the well-balanced property for realistic topographies, three hump shapes are considered with increasingly steeper bed slopes as shown in the top panels of Figure 9 (smooth on the left, steeper in the middle and rectangular on the right). For each hump shape, appropriate initial conditions are applied (see Table 3) with zero velocities to generate an unmoving free-surface elevation that leads to different wetting conditions around and/or at the humps.

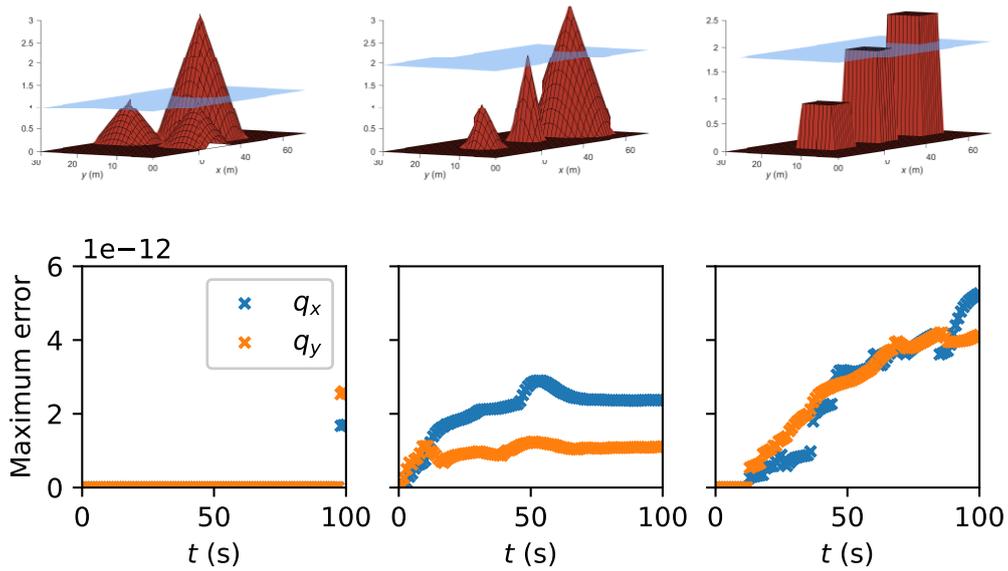

Figure 9: Verification of fidelity. Well-balanced property verification in the presence of wet-dry fronts with different levels of steepness in topography and different wetting conditions: smooth humps (left panels), steeper humps (middle panels) and rectangular humps (right panels). The top panels show the geometrical profiles of the humps in the domain area, and the lower panels include the time history of the maximum discharge errors where $q_x = hu$ and $q_y = hv$.

Table 3: Zero-velocity initial conditions applied to generate an unmoving free-surface flow for the three hump profiles shown in Figure 9.

| Hump profile | $h + z$ (m) | Wetting conditions | Reference |
|---|---|---|---|
| Smooth | 0.875 | Dry around the highest hump, critical ($h = 0$ m) over the two small humps | (Huang et al., 2013; Shirvani et al., 2021; Song et al., 2011) |
| Steeper | 1.78 | Dry around the highest hump, critical ($h = 0$ m) at the peak of the medium hump, wet above the shortest hump ($h > 0$ m) | (Kesserwani et al., 2018) |
| Rectangular | 1.95 | Dry around the highest hump, critical ($h = 0$ m) at the peak of the medium hump, wet above the shortest hump ($h > 0$ m) | (Kesserwani et al., 2018) |

GPU-HWFV1 simulations are run up to 100 s with a maximum refinement level $L = 8$ and an error threshold $\varepsilon = 10^{-3}$ (requiring around 3,000 timesteps to complete) while measuring the discharges. For GPU-HWFV1 to be deemed well-balanced, the free-surface elevation should stay undisturbed during the simulation, thus the discharges should not deviate from zero, within machine precision. The bottom panels of Figure 9 show the time histories of the maximum discharge errors (i.e., the maximum deviation from zero) for the three hump profiles. The errors are seen to become increasingly higher with increased irregularity in the hump profile, but nonetheless remain bounded as also observed for the CPU model counterparts (Kesserwani & Sharifian, 2020). This demonstrates that GPU-HWFV1 is well-balanced irrespective of the steepness of the bed slope and the presence of wet-dry zones and fronts in the domain area.

Next, GPU-HWFV1 is applied to reproduce a frictional dam-break flow ($n_M = 0.018$ m$^{1/3}$/s) for the smooth hump profile (top left panel, Figure 9). The initial dam-break flow conditions assume a water body of $h = 1.875$ m held by an imaginary dam located at $x = 16$ m with zero discharges. Using the same choice of $\varepsilon$ and $L$, a GPU-HWFV1 simulation is run up to 12 s. A GPU-FV1 simulation on the finest uniform grid accessible to GPU-HWFV1 is also performed to allow for like-for-like comparisons of flood depth profiles at outputs times reported in previous studies (Shirvani et al., 2021; Song et al., 2011). Figure 10 includes the 2D contour maps of the flood depths predicted by GPU-HWFV1 (left panel) compared to those predicted by GPU-FV1 (right panel) at 0, 6 and 12 s. At 0 s (top panel), both models are seen to start from the same flood depth profile. At 6 s (middle panel), both models predict that the small humps are completely submerged and that the dam-break wave has reached the large hump, and the $L^1$ error difference between depths predicted by GPU-HWFV1 and GPU-FV1 is $4.6 \times 10^{-4}$. There are similar wave patterns surrounding the large hump by 12 s (bottom panel) and the $L^1$ error is $9.2 \times 10^{-4}$. In all the predictions, GPU-HWFV1 shows symmetrical flood extent profiles that are similar to those reproduced GPU-FV1 and other hydrodynamic profiles reported in previous works (Shirvani et al., 2021; Song et al., 2011). This indicates that the GPU-HWFV1 implementation preserves the fidelity of well-established models used for real-world applications.

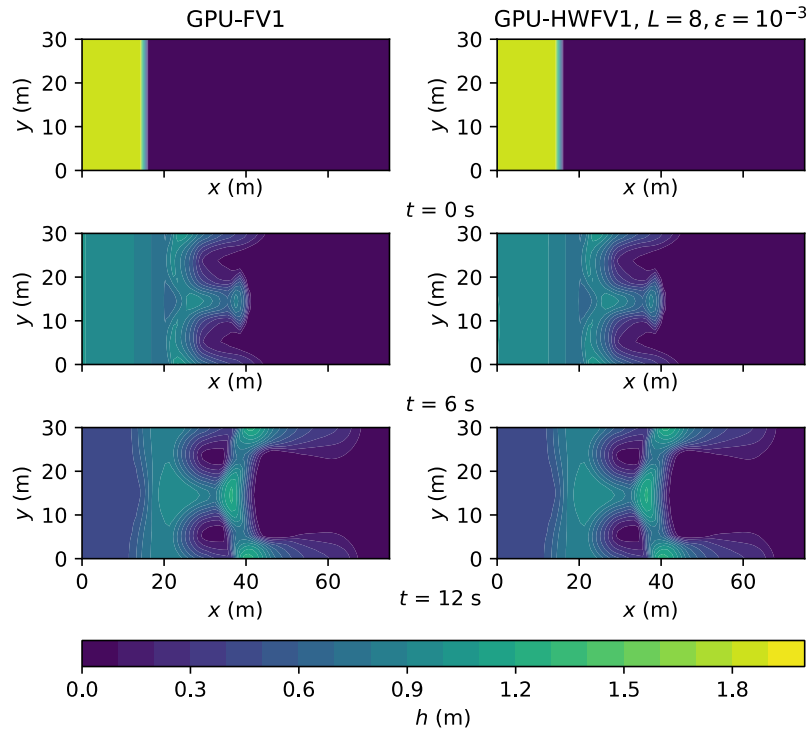

Figure 10: Verification of fidelity. Realistic dam-break flow with friction effects and moving wet-dry fronts. Flood depth profiles predicted by GPU-HWFV1 and GPU-FV1 on the left and right panels, respectively. At 0 s, the dam break wave emerges (top panels). At 6 s, the wave has submerged the small humps (middle panels). At 12 s, the wave starts to surround the large hump (bottom panels).

**3.2 Assessing runtime performance for synthetic test cases**

**Circular 2D dam-break flow.** This test case has often been used to verify new model implementations by capturing the symmetric propagation of shocks and rarefaction waves in the closed [-20 m, 20 m]$^2$ domain area (Toro, 2001). Initially, the water depth inside the cylindrical dam is 2.5 m, separating it from a water depth of 0.5 m elsewhere. The dam-break flow occurs over a frictionless and flat terrain, resulting in a shock moving radially outwards and a rarefaction wave moving radially inwards, which eventually collapses to form a secondary shock. It is first used to further verify GPU-HWFV1 using the same choice of $\varepsilon$ and $L$ as in Section 3.1 and by comparing its simulation outputs to those of CPU-HWFV1 and FV1-GPU. As in Kesserwani & Sharifian (2020), simulations are run up to $t = 3.5$ s for GPU-HWFV1, CPU-HWFV1 and FV1-GPU. Figure 11 shows the water depth centrelines predicted by the three models. GPU-HWFV1 predicts water depths that are identical to those predicted by CPU-HWFV1, GPU-FV1 and the benchmark solution. The benchmark solution was produced using the FV1 numerical solution to 1D radial form of the 2D shallow water equations using $256 \times 256$ cells, following (Toro, 2001).

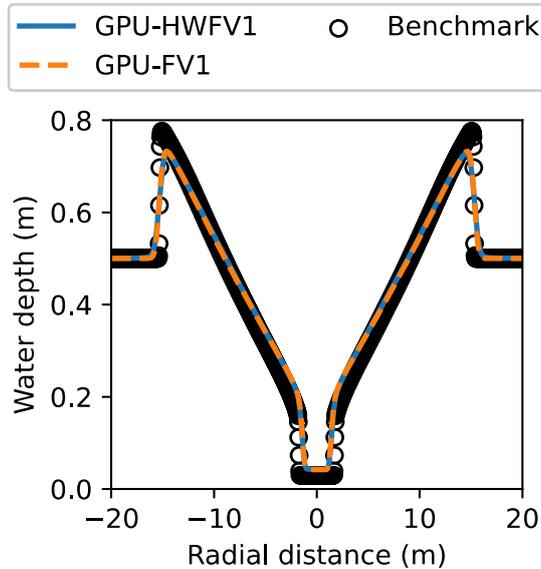

Figure 11: Circular 2D dam-break flow. Verification of GPU-HWFV1 using the same choice of $\varepsilon$ and $L$ as in Section 3.1 ($L = 8$ and $\varepsilon = 10^{-3}$): water depth centrelines at 3.5 s predicted by GPU-HWFV1, CPU-HWFV1, and GPU-FV1 compared to the benchmark solution.

To perform speed-up analysis, the models are rerun for the combinations of $\{\varepsilon, L\}$, and their runtimes were recorded for producing the speed-up ratios of GPU-HWFV1 relative to CPU-HWFV1 and GPU-FV1, respectively. Figure 12 contains the plots of the speed-up ratios with increasing maximum refinement level $L$, relative to CPU-HWFV1 in the left panel and to GPU-FV1 in the right panel. The black lines indicate the average speed-up ratios obtained for the three error thresholds and the dash-dotted lines indicate the breakeven point above which GPU-HWFV1 demonstrates speed-up (used also in the subsequent figures).

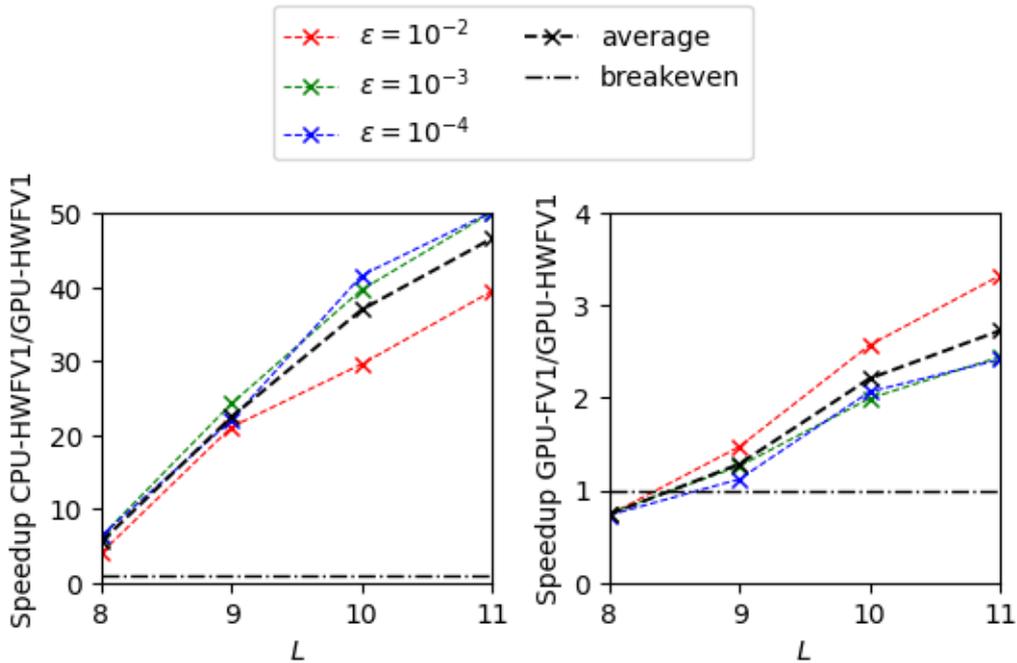

Figure 12: Circular 2D dam-break flow. Speed-up ratios to accomplish a 3.5 s simulation: GPU-HWFV1 over CPU-HWFV1 (left panel) and over GPU-FV1 (right panel).

GPU-HWFV1 is identified to be 5 to 46× faster than CPU-HWFV1. This speed-up is proportional to the increase in $L$ and decrease in $\varepsilon$. This suggests that wavelet-based grid adaptation is much more efficient when parallelised on the GPU, in particular as the maximum resolution refinement level is deepened and the sensitivity to refine resolution is increased. Compared to the runtime performance of FV1-GPU, GPU-HWFV1 is not faster in this test (right panel of Figure 12) until $L \geq 9$ for all the $\varepsilon$ values, reaching a maximum of 3× for the largest $\varepsilon = 10^{-2}$, and around 2× for the smaller $\varepsilon = 10^{-3}$ and $10^{-4}$. This means that GPU-HWFV1, despite the overhead costs from the MRA process, can still compete with the speed of a fine uniform-grid GPU-FV1 simulation even for a vigorous flow that would cause overrefinement on the adaptive grid. Namely, GPU-HWFV1 is likely to be faster than GPU-FV1 the deeper the grid resolution (which would lead to an excessively fine uniform grid for GPU-FV1) and the lower the sensitivity for triggering grid refinement. Next, a transient analysis of the speed-ups is performed in a longer simulation that sees a gradual change in the flow from vigorous to very smooth.

**Pseudo-2D dam-break flow.** This 1D dam-break flow test case has conventionally been used to verify shallow water models for a short simulation run (2.5 s) involving transient shock and rarefaction wave propagation in two opposite directions. It was used recently for a much longer simulation time (40 s) to assess speed-up for CPU-based adaptive grid models to their uniform grid counterparts by considering a flow with gradual transition from vigorous to smooth (Kesserwani et al., 2019; Kesserwani & Sharifian, 2020).

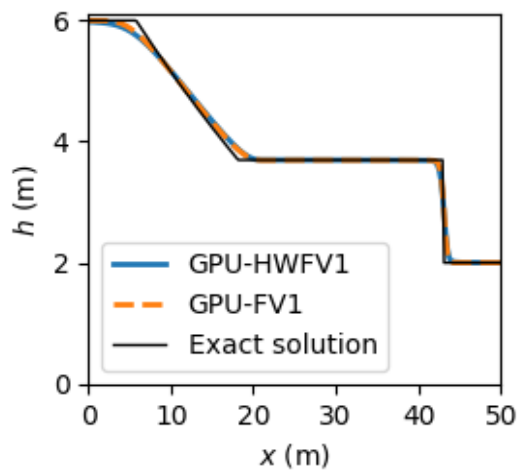

Figure 13: Pseudo-2D dam-break flow. Verification of GPU-HWFV1 using the same choice of $\varepsilon$ and $L$ as in Section 3.1 ($L = 8$ and $\varepsilon = 10^{-3}$): Water depths centerlines predicted by GPU-HWFV1, CPU-HWFV1 and GPU-FV1 at 2.5 s compared with the exact solution.

The domain area is 50 m × 25 m and assumed to be flat and frictionless with open boundary conditions. The dam, located at $x = 10$ m, initially separates an upstream water depth of 6 m from a downstream water depth of 2 m. After the dam removal, at $t = 0$ s, the shock and rarefaction waves remain present in the domain area up to 2.5 s. After 3 s, the shock wave has left the domain area from the downstream and the flow dynamics are only driven by the presence of the rarefaction wave until 10 s, after which it exits from the upstream. Therefore, after 10s, the flow dissipates gradually with increased smoothness until 40 s. The models are first verified by running simulations up to 2.5 s using the same choice of $\varepsilon$ and $L$ as in Section 3.1 ($L = 8$ and $\varepsilon = 10^{-3}$) for the HWFV1-based models and the finest uniform grid for the GPU-FV1 model. Figure 13 shows the plots of the water depth centrelines predicted by the models all showing a good agreement with the exact solution (Delestre et al., 2011).

To assess speed-up, GPU-HWFV1 and CPU-HWFV1 simulations are rerun for up to 40 s for the combinations of $\{\varepsilon, L\}$ alongside GPU-FV1 simulations on the finest uniform grid. Time histories of the runtimes are recorded throughout the 40 s simulations during which the flow transitions from vigorous to very smooth. Time series of the speed-up ratios of GPU-HWFV1 over CPU-HWFV1 and GPU-FV1 for the different values of $L$ and $\varepsilon$ are plotted in Figure 14.

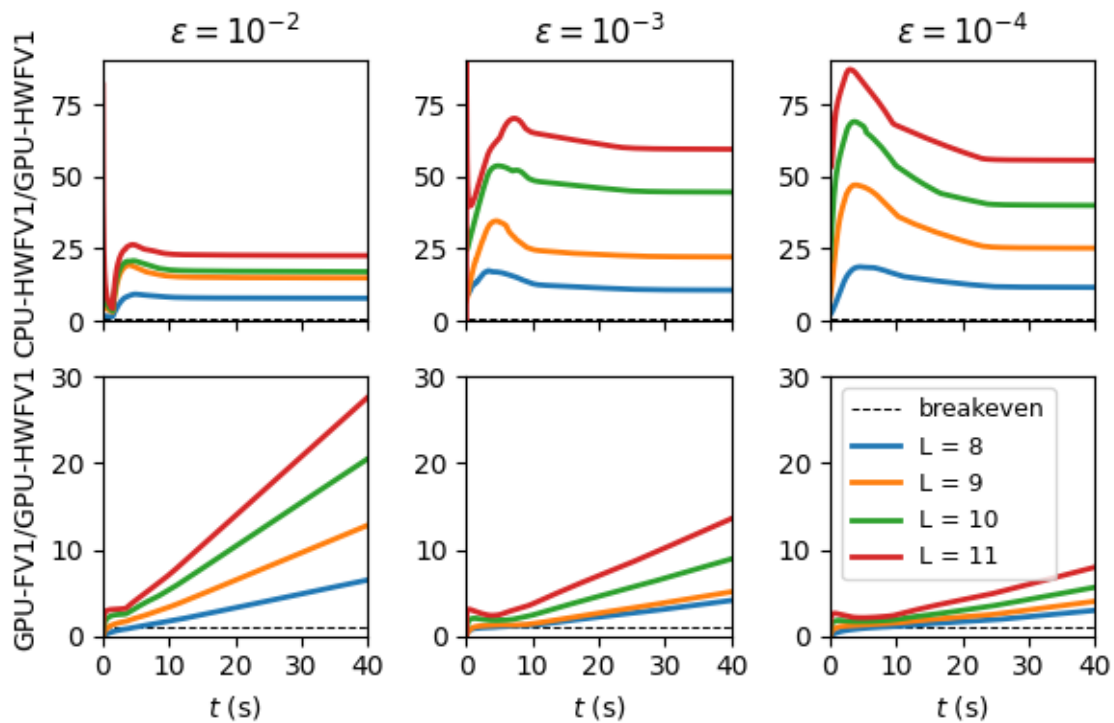

Figure 14: Pseudo-2D dam-break flow. Speed-up ratios of GPU-HWFV over CPU-HWFV1 (top panels), and over GPU-FV1 (bottom panels), for the three values of the threshold error $\varepsilon$ and considering different maximum refinement $L$.

Looking at the speed-up over CPU-HWFV1 (Figure 14, top panels), GPU-HWFV1 exceeds the breakeven for all but the largest $\varepsilon = 10^{-2}$ and the lowest maximum refinement level $L = 8$ up to 2.5 s (Figure 14, top left panel). This means that CPU-HWFV1 only remained as fast as GPU-HWFV1 when the flow included the shock and the rarefaction waves and for the setting with the least depth in resolution refinement and the least sensitivity to trigger grid refinement. However, even at $\varepsilon = 10^{-2}$, up to 8× speed-up is noted after 3 s when the shock wave is not present anymore. With any other combinations of $\{\varepsilon, L\}$, there is a significant demonstration of speed-up: with reduced $\varepsilon$ and increased $L$, GPU-HWFV1 becomes increasingly faster than CPU-HWFV1 up to reaching, for the highest $L$ and smallest $\varepsilon$ , an average speed-up of 68× throughout the simulation and a maximum speed-up of 88× at 2.5 s when flow discontinuities were still present. This confirms the benefit of parallelising the wavelet-based grid adaptation on the GPU as an alternative to the CPU version for general purpose modelling involving all types of flow.

In terms of speed-ups over GPU-FV1 (Figure 14, bottom panels), at $\varepsilon = 10^{-2}$, GPU-HWFV1 demonstrates a maximum speed-up of 25× when $L = 11$, though it could only outrun GPU-FV1 for $L \geq 9$, beyond which GPU-HWFV1 increasingly shows speed-up within increased smoothening in the flow. At $\varepsilon = 10^{-3}$, GPU-HWFV1's maximum speed-up reduces to 12× and outruns GPU-FV1 for $L \geq 10$, whereas for $L \leq 9$, it begins to demonstrate speed-up only after 10 s when the flow starts smoothening. This suggests to expect less speed-up over GPU-FV1 with increased sensitivity for triggering grid refinement with GPU-HWFV1 and reduced depth of the finest resolution. The same can be noted with $\varepsilon = 10^{-4}$, but here GPU-HWFV1 starts to be faster than FV1-GPU for $L \geq 9$ and the overall maximum speed-up reduces to 8× (reached again after 10 s when the flow is smoothening). These analyses indicate that an adaptive-grid GPU-HWFV1 simulation is likely to be more efficient than a uniform-grid GPU-FV1 simulation for very fine resolution modelling of gradual to smooth flows, with $L \geq 9$, and when the sensitivity to grid refinement is not maximal, with $\varepsilon > 10^{-4}$.

### 3.3 Further investigations into runtime performance: realistic flow simulations

**Dam-break wave interaction with an urban district.** This test case has widely been used for model verification (e.g. Caviedes-Voullième et al. (2020)) as it has a set of spatial experimental data for the water depth and the velocities (Soares-Frazão & Zech, 2008). It involves a dam-break wave propagation in a 36 m × 3.6 m smooth channel ($n_M = 0.01$) that includes a wall barrier with a gate initially separating an upstream water body of 0.4 m from a water depth of 0.011 m (Figure 15). Downstream of the gate, there are twenty-five 0.3 m × 0.3 m square blocks, with 0.1 m gaps. The ground height for the wall barrier and the square blocks is 2 m. Based on this height and the dimension reported in (Soares-Frazão & Zech, 2008), a DEM file was built at a resolution of 0.02 m × 0.02 m, made of 324,000 cells. The DEM includes the two rectangular blocks forming the wall barrier linked to the gate and the twenty-five square blocks. These discontinuous blocks are included in the grid and are accounted for as part of the well-balanced topography integration.

As the gate opens abruptly, a dam-break wave forms and flows swiftly to collide with the blocks. The blocks almost entirely impede the shock, creating a backwater zone upstream, while the unimpeded flow cascades through the gaps to form a hydraulic jump downstream as the simulation progresses (e.g. Fig. 24 in Soares-Frazão and Zech (2008)).

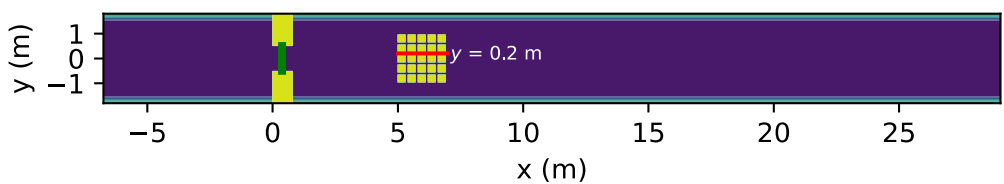

Figure 15: Dam-break wave interaction with an urban district. Top down view of the smooth channel with the gate indicated in green and topographic blocks coloured in yellow. Experimental depth and velocity data available along $y = 0.2$ m, indicated in red.

A 10 s simulation is run using GPU-HWFV1 with $L = 11$ for two values of $\varepsilon = \{10^{-4}, 10^{-3}\}$, and using GPU-FV1 on a uniform grid using the finest resolution accessible to GPU-HWFV1. Figure 16 shows the water depth (left panel) and velocity (right panel) profiles along $y = 0.2$ m at 6 s predicted by the GPU-HWFV1 and GPU-FV1 as well as the experimental profiles. All the models predicted profiles are within the expected range of agreement with the experimental profiles (Caviedes-Voullième et al., 2020; Kesserwani & Sharifian, 2020). Compared to the prediction made by GPU-FV1, those made by GPU-HWFV1 with $\varepsilon = 10^{-4}$ are closer than with $\varepsilon = 10^{-3}$ though the difference is not significant.

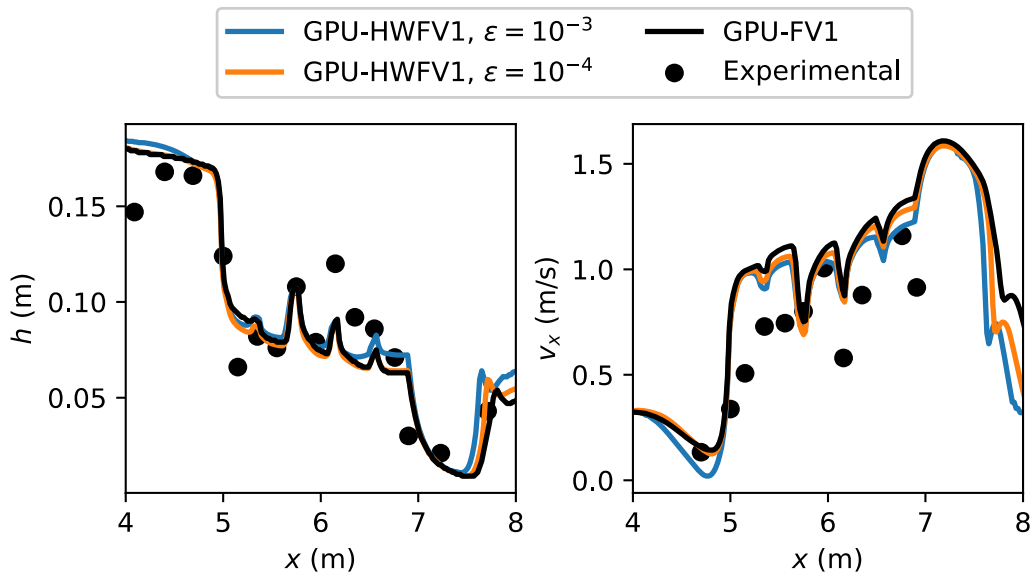

Figure 16: Dam-break wave interaction with an urban district. Depth (left panel) and velocity (right panel) profiles predicted along $y = 0.2$ m at 6 s by GPU-HWFV1 and GPU-FV1 compared to the experiments.

To analyse speed-ups, a CPU-HWFV1 simulation is also run. The recorded runtimes for the three models were used to calculate the time series of speed-up ratios of GPU-HWFV1 over CPU-HWFV1 and over GPU-FV1, which are plotted in the left and right panels of Figure 17. On average, GPU-HWFV1 is found 19× and 25× faster to run than CPU-HWFV1, with $\varepsilon = 10^{-3}$ and $10^{-4}$ respectively, throughout the 10 s simulation. Higher levels of speed-up are demonstrated with larger $\varepsilon$, which is in line with the findings in Section 3.2. GPU-HWFV1 is also faster than GPU-FV1 in this test, on average ~2.4× faster with both $\varepsilon = 10^{-3}$ and $10^{-4}$. This can be expected for a run with $L = 11$ accommodating the very fine resolution of the DEM. Up to 2 s, the run with GPU-HWFV1 at $\varepsilon = 10^{-3}$ demonstrates higher levels of speed-up than at $\varepsilon = 10^{-4}$, which is in line with the observations made in Section 3.2. In contrast, after 2 s, GPU-HWFV1 at $\varepsilon = 10^{-3}$ reduces the level of speed-up to become lower than with GPU-HWFV1 at $\varepsilon = 10^{-4}$. This could be due to GPU-HWFV1's higher sensitivity to grid refinement around the rectangular and square topographic blocks. Overall, GPU-HWFV1, besides being more performant than CPU-HWFV1, also remains faster than GPU-FV1 for this test. Supported by the analysis in Section 3.2, this can be expected given the maximised depth in the resolution level ($L = 11$) needed to accommodate the domain size to the fine resolution of the DEM.

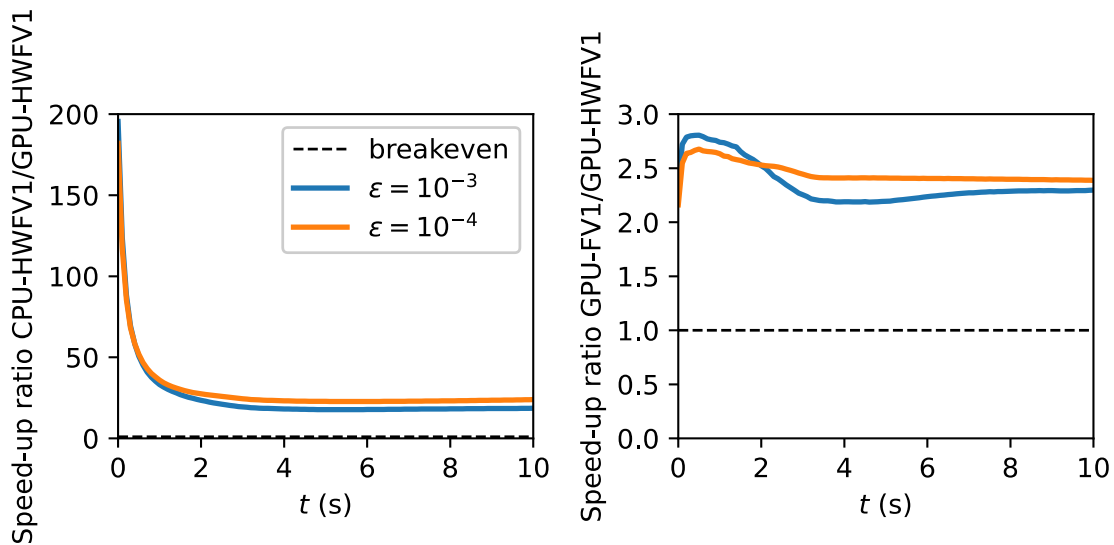

Figure 17: Dam-break wave interaction with an urban district. Speed-up ratios of GPU-HWFV1 over CPU-HWFV1 (left panel) and GPU-FV1 (right panel).

**Tsunami wave propagation over a complex beach.** The test case considers a 1:400 scaled replica of the 1993 Okushiri tsunami (Matsuyama & Tanaka, 2001). It has been used in other works for model verification and for runtime performance assessments of wavelet-based adaptive models versus their uniform counterparts for simulations on the CPU (Caviedes-Voullième et al., 2020; Kesserwani & Sharifian, 2020). It is here used to assess the runtime performance of the adaptive GPU-HWFV1 model versus CPU-HWFV1 and GPU-FV1 models.

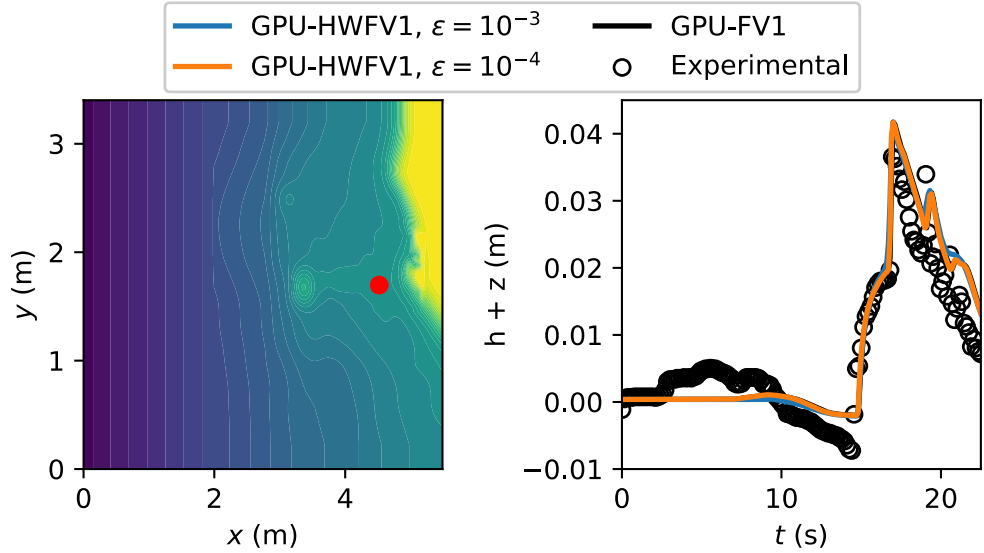

Figure 18: Tsunami wave propagation over a complex beach; (a) Topography contours over the domain area including the gauge point indicated in red. The tsunami-generated wave enters throughout the western boundary causing tsunami-generated flooding in the coastal area located in the eastern end (coloured in yellow); (b) Free-surface water elevation predicted by GPU-HWFV1 and GPU-FV1 compared to the experimental data.

The physical replica consists of a 5.488 m × 3.402 m smooth area ($n_M$ = 0.01 m$^{1/3}$/s) that has a uniform resolution of 0.014 m × 0.014 m on a DEM made of 163,840 cells (i.e. around twice fewer cells than the previous test case). The domain area has closed boundaries except for the western boundary through which a tsunami-generated inflow (Kesserwani & Sharifian, 2020) enters and eventually reaches the coastal area to the east, before which there is a gauge point ($x$ = 4.521 m, $y$ = 1.696 m) hit by the tsunami-generated flood wave. Experimental time histories of the free-surface water elevation are available at this point and will be used to verify the GPU-HWFV1 and GPU-FV1 models' ability to achieve a 22.5 s simulation. Figure 18a displays a view of the domain area including the gauge point location, marked by a red dot, and the coastal area at the eastern end (yellow colour). Given the smaller size of the domain area, the depth in the resolution level for the DEM to the domain size requires using $L$ = 9 in this test to run the GPU-HWFV1 simulations with $\varepsilon$ = {10$^{-3}$, 10$^{-4}$}. The GPU-FV1 simulation was run on a uniform grid at the DEM resolution. Figure 18b contains time histories of the free-surface water elevations predicted by the models, which are in a good agreement with the experimental time histories. It can be seen that all the models predict the expected gradual retraction in the free-surface elevation between 12 s and 15 s, followed by a sharp increase that peaks at around 17 s. GPU-HWFV1 at $\varepsilon = 10^{-4}$ leads to predictions that are visually indistinguishable from those predicted by GPU-FV1. With $\varepsilon = 10^{-3}$, the predictions remain comparable subject to small, localised discrepancies at times where there is a sharp flow transition such as at around 18 and 21 s.

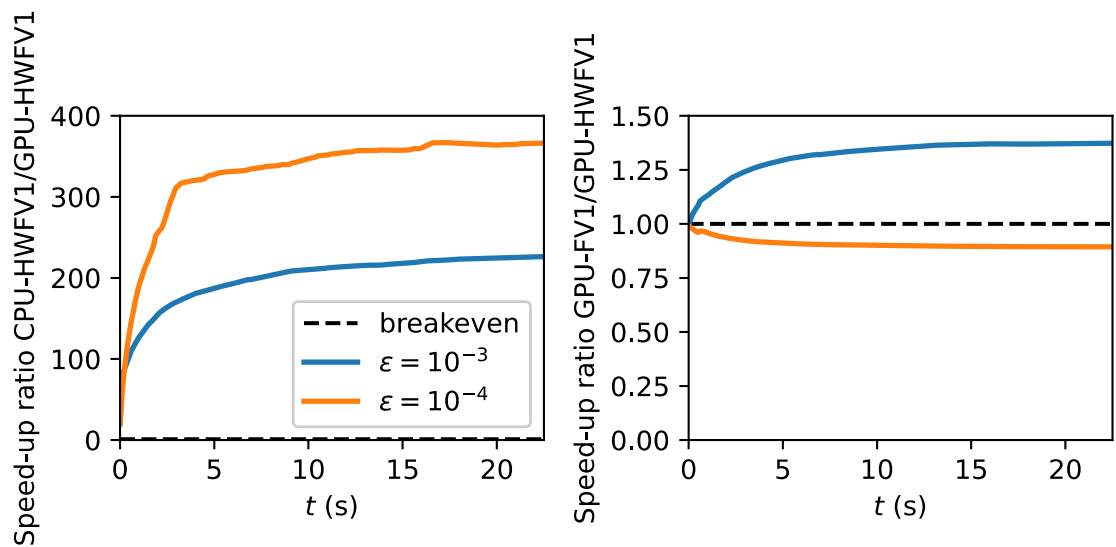

Figure 19: Tsunami wave propagation over a complex beach. Speed-up ratios of GPU-HWFV1 over CPU-HWFV1 (left panel) and over GPU-FV1 (right panel).

Figure 19 contains the plots of the time histories of the speed-up ratios for GPU-HWFV1 over CPU-HWFV1 run with similar setting (left panel) and over GPU-FV1 (right panel) during the 22.5 s simulation. GPU-HWFV1 is seen to be significantly faster than CPU-HWFV1, leading to average speed-ups of 200× and 400× with $\varepsilon = 10^{-3}$ and $10^{-4}$, respectively. Compared to the previous test case, the terrain is complex all over the domain and the grid cannot be coarsened much, leading to an overrefined grid that is further refined by flow disturbances during HWFV1 simulations. In such a case, GPU-HWFV1 demonstrates remarkable speed-up over CPU-HWFV1, which increases as $\varepsilon$ is decreased from $10^{-3}$ to $\varepsilon = 10^{-4}$ in line with the observations in Section 3.2, but the speed-up doubles in this test.

Compared to GPU-FV1, GPU-HWFV1 only demonstrates speed-up with $\varepsilon = 10^{-3}$ (around 1.25×) and is slightly slower to run with $\varepsilon = 10^{-4}$ where its speed-up falls below the breakeven line. This implies that it is worthwhile to perform wavelet-based grid adaptation in this test for $\varepsilon = 10^{-3}$ but not for $\varepsilon = 10^{-4}$, at which the grid is overrefined (due to the terrain and flow disturbances). Overall, GPU-HWFV1 is shown to be generally faster than CPU-HWFV1 but could not outrun GPU-FV1 for $\varepsilon = 10^{-4}$ (which leads to an overrefined grid) and for $L = 9$ (to accommodate a relatively small domain). Nonetheless, for $\varepsilon = 10^{-3}$, GPU-HWFV1 remains a viable choice over GPU-FV1.

## 4. Conclusion

This paper presented an adaptive first-order finite volume (FV1) shallow water model with a dynamic-in-time wavelet-based grid adaptation process driven by the "multiresolution analysis" (MRA) of the Haar wavelet (HW) that is fully parallelised on the graphics processing unit (GPU), called GPU-HWFV1. The MRA involves a nested hierarchy of grids, where the coarsest grid in the hierarchy is made up of a single cell while the finest grid is made up of $2^L \times 2^L$ cells, where $L$ is a user-specified maximum refinement level. In the "encoding" (coarsening) process, a user-specified error threshold $\varepsilon$ is needed to flag significant "details" to decide which cells to include in the non-uniform grid. The encoding process results in a tree-like structure of significant details that is traversed in the "decoding" (refinement) process by applying a sequential depth-first traversal (DFT) algorithm to identify the cells making up the non-uniform grid. Encoding and decoding have been parallelised on the GPU by adopting the indexing of a Z-order space-filling curve to ensure coalesced memory access. Meanwhile, the DFT algorithm has been replaced with a parallel tree traversal (PTT) algorithm to traverse the tree of significant details on the GPU with minimal warp divergence. The PTT algorithm also allows to easily identify the neighbour cells of each cell in the non-uniform grid for performing the FV1 update in parallel.

GPU-HWFV1 was first verified and then its runtime performance was assessed against a sequential predecessor running on the central processing unit (CPU-HWFV1) as well as the reference uniform-grid FV1 counterpart parallelised on the GPU (GPU-FV1) ran on the finest grid accessible to the HWFV1 models. The verification was performed using $\varepsilon = 10^{-3}$ (recommended for flood modelling) and $L = 8$ for four synthetic test cases involving motionless, vigorous, gradual and smooth flows. A systematic runtime performance assessment was performed for two synthetic test cases involving dam-break flow, where a lower and higher bound for $\varepsilon = \{10^{-2}, 10^{-3}, 10^{-4}\}$ were also considered alongside an increase in the maximum refinement level $L = \{8, 9, 10, 11\}$. Verification and runtime performance assessments were finally performed for realistic test cases with digital elevation models (DEMs) for which the value of $L$ was chosen to match the DEM resolution, while GPU-HWFV1 was run with $\varepsilon = \{10^{-3}, 10^{-4}\}$.

GPU-HWFV1's overall performance for all the test cases provided strong evidence that it delivers a similar level of fidelity as GPU-FV1 in replicating the realistic flows including the presence of uneven topographies, wet-dry fronts and friction effects. In terms of runtime performance over CPU-HWFV1, GPU-HWFV1 yielded significant speed-ups for all the test cases, ranging between 20× to 400×. Hence, this work offers compelling evidence for porting the GPU-parallelised wavelet-based grid adaptation process to FV1 models in other fields. From the systematic runtime performance assessment for the synthetic test cases, GPU-HWFV1 tends to demonstrate speed-up of around 1.1× to 30× over GPU-FV1 for $L \geq 9$ and/or by avoiding the smallest $\varepsilon = 10^{-4}$. For the realistic test cases, GPU-HWFV1 showed speed-up over GPU-FV1 for the test with $L = 11$ and for $\varepsilon = 10^{-3}$ for the test with $L = 9$. Hence, GPU-HWFV1 can be favoured to gain runtime performance over GPU-FV1 for shallow water modelling over real DEMs, namely with an increased fineness in the DEM resolution and an increased domain size.

## References

Arpaia, L., & Ricchiuto, M. (2018). r−adaptation for Shallow Water flows: conservation, well balancedness, efficiency. *Computers & Fluids*, *160*, 175–203. https://doi.org/10.1016/J.COMPFLUID.2017.10.026

Bader, M. (2013). *Space-Filling Curves* (1st ed.). Springer Berlin, Heidelberg.

Beckingsale, D., Gaudin, W., Herdman, A., & Jarvis, S. (2015). Resident Block-Structured Adaptive Mesh Refinement on Thousands of Graphics Processing Units. *2015 44th International Conference on Parallel Processing*, 61–70. https://doi.org/10.1109/ICPP.2015.15

Bédorf, J., Gaburov, E., & Portegies Zwart, S. (2012). A sparse octree gravitational N-body code that runs entirely on the GPU processor. *Journal of Computational Physics*, *231*(7), 2825–2839. https://doi.org/https://doi.org/10.1016/j.jcp.2011.12.024

Berger, M. J., & Colella, P. (1989). Local adaptive mesh refinement for shock hydrodynamics. *Journal of Computational Physics*, *82*(1), 64–84. https://doi.org/https://doi.org/10.1016/0021-9991(89)90035-1

Brix, K., Melian, S. S., Müller, S., & Schieffer, G. (2009). Parallelisation of Multiscale-Based Grid Adaptation using Space-Filling Curves. *Esaim: Proceedings*, *29*, 108–129.

Brodtkorb, A. R., Hagen, T. R., & Sætra, M. L. (2013). Graphics processing unit (GPU) programming strategies and trends in GPU computing. *Journal of Parallel and Distributed Computing*, *73*(1), 4–13. https://doi.org/https://doi.org/10.1016/j.jpdc.2012.04.003

Buttinger-Kreuzhuber, A., Konev, A., Horváth, Z., Cornel, D., Schwerdorf, I., Blöschl, G., & Waser, J. (2022). An integrated GPU-accelerated modeling framework for high-resolution simulations of rural and urban flash floods. *Environmental Modelling & Software*, *156*, 105480. https://doi.org/https://doi.org/10.1016/j.envsoft.2022.105480

Carlotto, T., Borges Chaffe, P. L., Innocente dos Santos, C., & Lee, S. (2021). SW2D-GPU: A two-dimensional shallow water model accelerated by GPGPU. *Environmental Modelling & Software*, *145*, 105205. https://doi.org/https://doi.org/10.1016/j.envsoft.2021.105205

Caviedes-Voullième, D., Gerhard, N., Sikstel, A., & Müller, S. (2020). Multiwavelet-based mesh adaptivity with Discontinuous Galerkin schemes: Exploring 2D shallow water problems. *Advances in Water Resources*, *138*, 103559. https://doi.org/10.1016/J.ADVWATRES.2020.103559

Caviedes-Voullième, D., & Kesserwani, G. (2015). Benchmarking a multiresolution discontinuous Galerkin shallow water model: Implications for computational hydraulics. *Advances in Water Resources*, *86*, 14–31. https://doi.org/10.1016/J.ADVWATRES.2015.09.016

Caviedes-Voullième, D., Morales-Hernández, M., Norman, M. R., & Özgen-Xian, I. (2023). SERGHEI (SERGHEI-SWE) v1.0: a performance-portable high-performance parallel-computing shallow-water solver for hydrology and environmental hydraulics. *Geoscientific Model Development*, *16*(3), 977–1008. https://doi.org/10.5194/gmd-16-977-2023

Chitalu, F. M., Dubach, C., & Komura, T. (2018). Bulk-Synchronous Parallel Simultaneous BVH Traversal for Collision Detection on GPUs. *Proceedings of the ACM SIGGRAPH Symposium on Interactive 3D Graphics and Games.* https://doi.org/10.1145/3190834.3190848

Dazzi, S., Vacondio, R., & Mignosa, P. (2020). Internal boundary conditions for a GPU-accelerated 2D shallow water model: Implementation and applications. *Advances in Water Resources*, *137*, 103525. https://doi.org/https://doi.org/10.1016/j.advwatres.2020.103525

Deiterding, R., Domingues, M. O., & Schneider, K. (2020). Multiresolution analysis as a criterion for effective dynamic mesh adaptation – A case study for Euler equations in the SAMR framework AMROC. *Computers & Fluids*, *205*, 104583. https://doi.org/https://doi.org/10.1016/j.compfluid.2020.104583

de la Asunción, M., & Castro, M. J. (2017). Simulation of tsunamis generated by landslides using adaptive mesh refinement on GPU. *Journal of Computational Physics*, *345*, 91–110. https://doi.org/10.1016/J.JCP.2017.05.016

Delmas, V., & Soulaïmani, A. (2022). Multi-GPU implementation of a time-explicit finite volume solver using CUDA and a CUDA-Aware version of OpenMPI with application to shallow water flows. *Computer Physics Communications*, *271*, 108190. https://doi.org/https://doi.org/10.1016/j.cpc.2021.108190

Domingues, M. O., Deiterding, R., Moreira Lopes, M., Gomes, A. K. F., Mendes, O., & Schneider, K. (2019). Wavelet-based parallel dynamic mesh adaptation for magnetohydrodynamics in the AMROC framework. *Computers & Fluids*, *190*, 374–381. https://doi.org/https://doi.org/10.1016/j.compfluid.2019.06.025

Donat, R., Martí, M. C., Martínez-Gavara, A., & Mulet, P. (2014). Well-Balanced Adaptive Mesh Refinement for shallow water flows. *Journal of Computational Physics*, *257*, 937–953. https://doi.org/10.1016/J.JCP.2013.09.032

Dunning, D., Marts, W., Robey, R. W., & Bridges, P. (2020). Adaptive mesh refinement in the fast lane. *Journal of Computational Physics*, *406*, 109193. https://doi.org/https://doi.org/10.1016/j.jcp.2019.109193

Ferrari, A., Vacondio, R., & Mignosa, P. (2023). High-resolution 2D shallow water modelling of dam failure floods for emergency action plans. *Journal of Hydrology*, *618*, 129192. https://doi.org/https://doi.org/10.1016/j.jhydrol.2023.129192

Flood Modeller 2D. (2022). *Flood Modeller by Jacobs.*

Gerhard, N., Caviedes-Voullième, D., Müller, S., & Kesserwani, G. (2015). Multiwavelet-based grid adaptation with discontinuous Galerkin schemes for shallow water equations. *Journal of Computational Physics*, *301*, 265–288. https://doi.org/10.1016/J.JCP.2015.08.030

Ghazizadeh, M. A., Mohammadian, A., & Kurganov, A. (2020). An adaptive well-balanced positivity preserving central-upwind scheme on quadtree grids for shallow water equations. *Computers & Fluids*, *208*, 104633. https://doi.org/https://doi.org/10.1016/j.compfluid.2020.104633

Gillis, T., & van Rees, W. M. (2022). MURPHY—A Scalable Multiresolution Framework for Scientific Computing on 3D Block-Structured Collocated Grids. *SIAM Journal on Scientific Computing*, *44*(5), C367–C398. https://doi.org/10.1137/21M141676X

Giuliani, A., & Krivodonova, L. (2019). Adaptive mesh refinement on graphics processing units for applications in gas dynamics. *Journal of Computational Physics*, *381*, 67–90. https://doi.org/https://doi.org/10.1016/j.jcp.2018.12.019

Goldfarb, M., Jo, Y., & Kulkarni, M. (2013). General Transformations for GPU Execution of Tree Traversals. *Proceedings of the International Conference on High Performance Computing, Networking, Storage and Analysis*. https://doi.org/10.1145/2503210.2503223

Gong, W., Shimizu, Y., & Iwasaki, T. (2020). A Case Study of Flood Modeling with Adaptive Mesh Refinement. *Proceedings of the 22nd IAHR APD Congress (Sapporo, 2020)*.

Gordillo, G., Morales-Hernández, M., Echeverribar, I., Fernández-Pato, J., & García-Navarro, P. (2020). A GPU-based 2D shallow water quality model. *Journal of Hydroinformatics*, *22*(5), 1182–1197. https://doi.org/10.2166/hydro.2020.030

Greenberg, J. M., & Leroux, A. Y. (1996). A Well-Balanced Scheme for the Numerical Processing of Source Terms in Hyperbolic Equations. *SIAM Journal on Numerical Analysis*, *33*(1), 1–16. https://doi.org/10.1137/0733001

Haleem, D. A., Kesserwani, G., & Caviedes-Voullième, D. (2015). Haar wavelet-based adaptive finite volume shallow water solver. *Journal of Hydroinformatics*, *17*(6), 857–873. https://doi.org/10.2166/hydro.2015.039

Han, H., Hou, J., Xu, Z., Jing, H., Gong, J., Zuo, D., Li, B., Yang, S., Kang, Y., & Wang, R. (2022). A GPU-Accelerated Hydrodynamic Model for Urban Rainstorm Inundation Simulation: A Case Study in China. *KSCE Journal of Civil Engineering*, *26*(3), 1494–1504. https://doi.org/10.1007/s12205-021-2158-3

Holzbecher, E. (2022). Adaptive Mesh Refinement for Dam-Break Models using the Shallow Water Equations . *Journal of the Civil Engineering Forum*, *9*(1), 79–90. https://doi.org/10.22146/jcef.4260

Hou, J., Liang, Q., Zhang, H., & Hinkelmann, R. (2015). An efficient unstructured MUSCL scheme for solving the 2D shallow water equations. *Environmental Modelling & Software*, *66*, 131–152. https://doi.org/10.1016/J.ENVSOFT.2014.12.007

Huang, Y., Zhang, N., & Pei, Y. (2013). Well-Balanced Finite Volume Scheme for Shallow Water Flooding and Drying Over Arbitrary Topography. *Engineering Applications of Computational Fluid Mechanics*, *7*(1), 40–54. https://doi.org/10.1080/19942060.2013.11015452

Hu, R., Fang, F., Salinas, P., & Pain, C. C. (2018). Unstructured mesh adaptivity for urban flooding modelling. *Journal of Hydrology*, *560*, 354–363. https://doi.org/https://doi.org/10.1016/j.jhydrol.2018.02.078

InfoWorks ICM. (2018). *GPU Runtime Results for 2D InfoWorks ICM Models*.

Jeong, W., Yoon, J. S., & Cho, Y. S. (2012). Numerical study on effects of building groups on dam-break flow in urban areas. *Journal of Hydro-Environment Research*, *6*(2), 91–99. https://doi.org/10.1016/J.JHER.2012.01.001

Julius, E. T. K., & Marie, F. (2021). A Wavelet-Adaptive Method for Multiscale Simulation of Turbulent Flows in Flying Insects. *Communications in Computational Physics*, *30*(4), 1118–1149. https://doi.org/https://doi.org/10.4208/cicp.OA-2020-0246

Karras, T. (2012, November 26). *Thinking Parallel, Part II: Tree Traversal on the GPU*.

Kesserwani, G., Ayog, J. L., & Bau, D. (2018). Discontinuous Galerkin formulation for 2D hydrodynamic modelling: Trade-offs between theoretical complexity and practical convenience. *Computer Methods in Applied Mechanics and Engineering*, *342*, 710–741. https://doi.org/10.1016/J.CMA.2018.08.003

Kesserwani, G., & Sharifian, M. K. (2020). (Multi)wavelets increase both accuracy and efficiency of standard Godunov-type hydrodynamic models: Robust 2D approaches. *Advances in Water Resources*, *144*, 103693. https://doi.org/10.1016/J.ADVWATRES.2020.103693

Kesserwani, G., Shaw, J., Sharifian, M. K., Bau, D., Keylock, C. J., Bates, P. D., & Ryan, J. K. (2019). (Multi)wavelets increase both accuracy and efficiency of standard Godunov-type hydrodynamic models. *Advances in Water Resources*, *129*, 31–55. https://doi.org/10.1016/J.ADVWATRES.2019.04.019

Kévin, P., Golay, F., & Marcer, R. (2017). *Adaptive Mesh Refinement Method Applied to Shallow Water Model: A Mass Conservative Projection*. https://doi.org/10.14311/TPFM.2017.032

Lakhlifi, Y., Daoudi, S., & Boushaba, F. (2018). Dam-Break Computations by a Dynamical Adaptive Finite Volume Method. *Journal of Applied Fluid Mechanics*, *11*(6), 1543–1556. https://doi.org/10.29252/jafm.11.06.28564

Liang, Q. (2010). Flood Simulation Using a Well-Balanced Shallow Flow Model. *Journal of Hydraulic Engineering*, *136*(9), 669–675. https://doi.org/10.1061/(ASCE)HY.1943-7900.0000219

Liang, Q., Hou, J., & Xia, X. (2015). Contradiction between the C-property and mass conservation in adaptive grid based shallow flow models: cause and solution. *International Journal for Numerical Methods in Fluids*, *78*(1), 17–36. https://doi.org/https://doi.org/10.1002/fld.4005

Lohr, C. (2009). *GPU-Based Parallel Stackless BVH Traversal for Animated Distributed Ray Tracing*.

Lung, K., Brown-Dymkoski, E., Guerrero, V., Doran, E., Museth, K., Balme, J., Urberger, B., Kessler, A., Jones, S., Moses, B., & Crognale, A. (2016). Efficient Combustion Simulation via the Adaptive Wavelet Collocation Method. *APS March Meeting 2016*.

Merrill, D. (2022). *CUB software package*. https://nvlabs.github.io/cub/

MIKE 21 GPU. (2019). *MIKE Powered by DHI: GPU - Guidelines*.

Nam, M., Kim, J., & Nam, B. (2016). Parallel Tree Traversal for Nearest Neighbor Query on the GPU. *2016 45th International Conference on Parallel Processing (ICPP)*, 113–122. https://doi.org/10.1109/ICPP.2016.20

NVIDIA. (2023). *CUDA C++ Programming Guide*. https://docs.nvidia.com/cuda/cuda-c-programming-guide/

Qin, X., LeVeque, R. J., & Motley, M. R. (2019). Accelerating an Adaptive Mesh Refinement Code for Depth-Averaged Flows Using GPUs. *Journal of Advances in Modeling Earth Systems*, *11*(8), 2606–2628. https://doi.org/https://doi.org/10.1029/2019MS001635

Sætra, M., Brodtkorb, A., & Lie, K.-A. (2014). Efficient GPU-Implementation of Adaptive Mesh Refinement for the Shallow-Water Equations. *Journal of Scientific Computing*, *63*. https://doi.org/10.1007/s10915-014-9883-4

Sagan, H. (1994). *Space-Filling Curves* (1st ed.). Springer New York, NY.

Sanz-Ramos, M., López-Gómez, D., Bladé, E., & Dehghan-Souraki, D. (2023). A CUDA Fortran GPU-parallelised hydrodynamic tool for high-resolution and long-term eco-hydraulic modelling. *Environmental Modelling & Software*, *161*, 105628. https://doi.org/https://doi.org/10.1016/j.envsoft.2023.105628

Schive, H.-Y., ZuHone, J. A., Goldbaum, N. J., Turk, M. J., Gaspari, M., & Cheng, C.-Y. (2018). gamer-2: a GPU-accelerated adaptive mesh refinement code – accuracy, performance, and scalability. *Monthly Notices of the Royal Astronomical Society*, *481*(4), 4815–4840. https://doi.org/10.1093/mnras/sty2586

Sedgewick, R., & Wayne, K. (2011). *Algorithms, 4th Edition*. Addison-Wesley.

Semakin, A. N., & Rastigejev, Y. (2020). Optimized wavelet-based adaptive mesh refinement algorithm for numerical modeling of three-dimensional global-scale atmospheric chemical transport. *Quarterly Journal of the Royal Meteorological Society*, *146*(729), 1564–1574. https://doi.org/https://doi.org/10.1002/qj.3752

Shaw, J., Kesserwani, G., Neal, J., Bates, P., & Sharifian, M. K. (2021). LISFLOOD-FP 8.0: the new discontinuous Galerkin shallow-water solver for multi-core CPUs and GPUs. *Geosci. Model Dev.*, *14*(6), 3577–3602. https://doi.org/10.5194/gmd-14-3577-2021

Shirvani, M., Kesserwani, G., & Richmond, P. (2021). Agent-based simulator of dynamic flood-people interactions. *Journal of Flood Risk Management*, *14*(2), e12695. https://doi.org/https://doi.org/10.1111/jfr3.12695

Song, L., Zhou, J., Li, Q., Yang, X., & Zhang, Y. (2011). An unstructured finite volume model for dam-break floods with wet/dry fronts over complex topography. *International Journal for Numerical Methods in Fluids*, *67*(8), 960–980. https://doi.org/https://doi.org/10.1002/fld.2397

Soni, V., Hadjadj, A., Roussel, O., & Moebs, G. (2019). Parallel multi-core and multi-processor methods on point-value multiresolution algorithms for hyperbolic conservation laws. *Journal of Parallel and Distributed Computing*, *123*, 192–203. https://doi.org/https://doi.org/10.1016/j.jpdc.2018.09.016

TUFLOW HPC. (2018). *TUFLOW Classic/HPC User Manual*.

Wahib, M., Maruyama, N., & Aoki, T. (2016). Daino: A High-Level Framework for Parallel and Efficient AMR on GPUs. *SC '16: Proceedings of the International Conference for High Performance Computing, Networking, Storage and Analysis*, 621–632. https://doi.org/10.1109/SC.2016.52

Wallwork, J. G., Barral, N., Kramer, S. C., Ham, D. A., & Piggott, M. D. (2020). Goal-oriented error estimation and mesh adaptation for shallow water modelling. *SN Applied Sciences*, *2*(6), 1053. https://doi.org/10.1007/s42452-020-2745-9

Wang, Y., Liang, Q., Kesserwani, G., & Hall, J. W. (2011). A 2D shallow flow model for practical dam-break simulations. *Journal of Hydraulic Research*, *49*(3), 307–316. https://doi.org/10.1080/00221686.2011.566248

Xia, X., Liang, Q., & Ming, X. (2019). A full-scale fluvial flood modelling framework based on a high-performance integrated hydrodynamic modelling system (HiPIMS). *Advances in Water Resources*, *132*, 103392. https://doi.org/10.1016/J.ADVWATRES.2019.103392

Zeidan, D., Schmidt, A. A., Kozakevicius, A. J., & Jakobsson, S. (2022). Towards parallel WENO wavelet methods for the simulation of compressible two-fluid models. *AIP Conference Proceedings*, *2425*(1), 20017. https://doi.org/10.1063/5.0082038

Zhang, M., Huang, W., & Qiu, J. (2021). A High-Order Well-Balanced Positivity-Preserving Moving Mesh DG Method for the Shallow Water Equations With Non-Flat Bottom Topography. *Journal of Scientific Computing*, *87*(3), 88. https://doi.org/10.1007/s10915-021-01490-3

Zhou, F., Chen, G., Huang, Y., Yang, J. Z., & Feng, H. (2013). An adaptive moving finite volume scheme for modeling flood inundation over dry and complex topography. *Water Resources Research*, *49*(4), 1914–1928. https://doi.org/https://doi.org/10.1002/wrcr.20179

Zola, W. M. N., Bona, L. C. E., & Silva, F. (2014). Fast GPU parallel N-Body tree traversal with Simulated Wide-Warp. *2014 20th IEEE International Conference on Parallel and Distributed Systems (ICPADS)*, 718–725. https://doi.org/10.1109/PADSW.2014.7097874

**Figures**

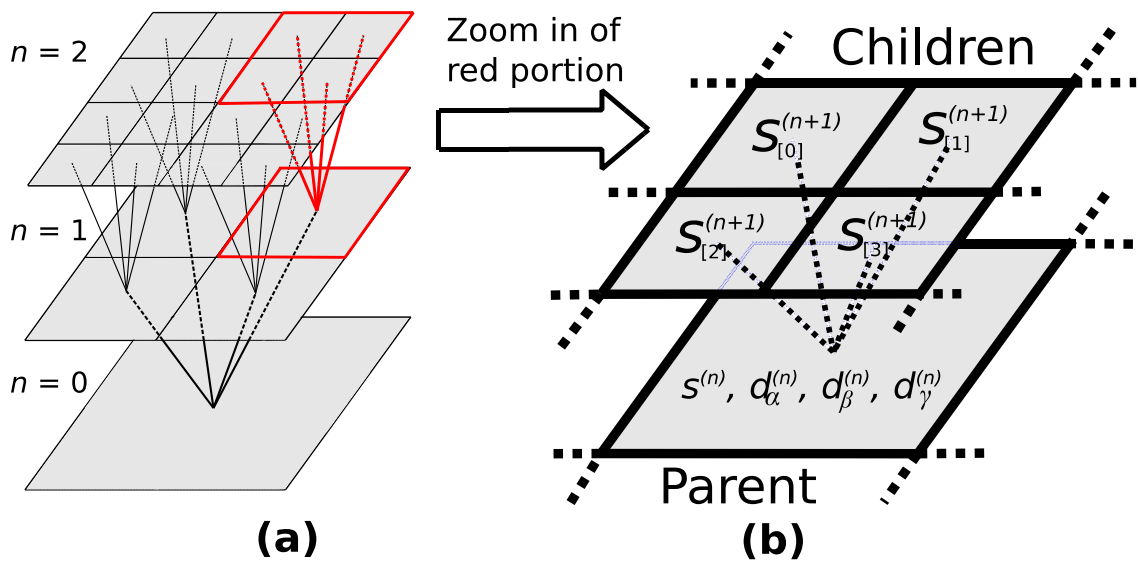

Figure 1: Multiresolution analysis (MRA). Left panel shows a hierarchy of grids involved in the MRA, with a maximum refinement level $L = 2$. Right panel shows how four cells at refinement level $n + 1$, called "children", are related to a single cell at refinement level $n$, called "parent". Also shown are the coefficients $s$ and details $d$ that are involved in computations to realise the MRA.

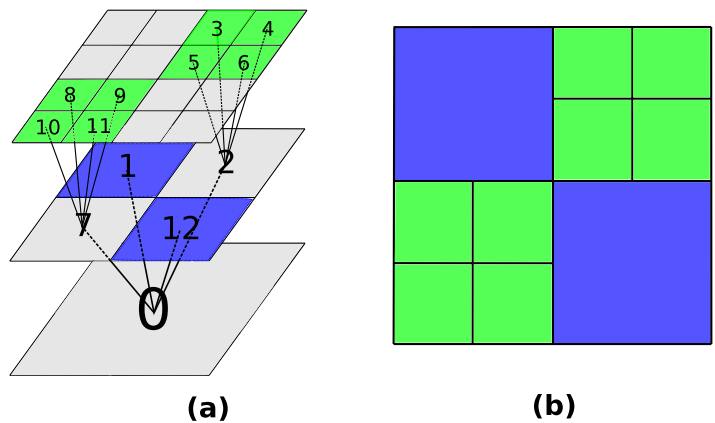

Figure 2: Left panel shows the tree-like structure obtained after flagging significant details during the process of encoding; the cells where the tree terminates are called "leaf" cells (highlighted in green and blue). Right panel shows how the leaf cells are assembled into a non-uniform grid.

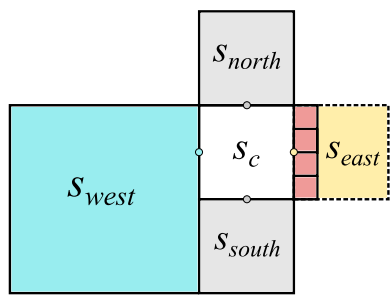

Figure 3: Finding the neighbours of a cell in a non-uniform grid to retrieve the sets $S_c$, $S_{west}$, $S_{east}$, $S_{north}$ and $S_{south}$ in order to compute the spatial operator $\mathbf{L}_c$.

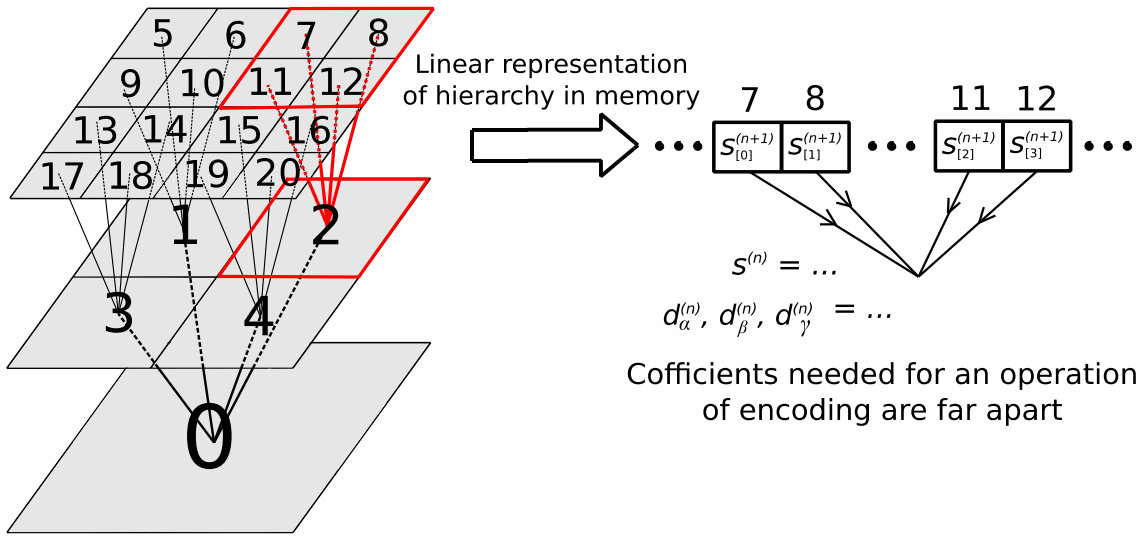

Figure 4: Indexing the hierarchy of grids in row-major order (left panel) and the corresponding locations of the children $s^{(n+1)}_{[0]}, s^{(n+1)}_{[1]}, s^{(n+1)}_{[2]}$ and $s^{(n+1)}_{[3]}$ in memory (right panel).

| | $i$ | | | |
|---|---|---|---|---|
| $j$ | 00 | 01 | 10 | 11 |
| 00 | 0000 | 0001 | 0100 | 0101 |
| 01 | 0010 | 0011 | 0110 | 0111 |
| 10 | 1000 | 1001 | 1100 | 1101 |
| 11 | 1010 | 1011 | 1110 | 1111 |

**(a)**

| | $i$ | | | |
|---|---|---|---|---|
| $j$ | 0 | 1 | 2 | 3 |
| 0 | 0 | 1 | 4 | 5 |
| 1 | 2 | 3 | 6 | 7 |
| 2 | 8 | 9 | 12 | 13 |
| 3 | 10 | 11 | 14 | 15 |

**(b)**

Figure 5: Creation of a Z-order curve for a $2^2 \times 2^2$ grid. The left panel shows how the binary forms of the $i$ and $j$ indices of each cell making up a $2^2 \times 2^2$ grid are bit interleaved (alternating red and black digits) to yield so-called Morton codes (also in binary). The right panel shows how these Morton codes (in decimal form) are followed in ascending order to create a Z-order curve and enforce Z-order indexing of the grid.

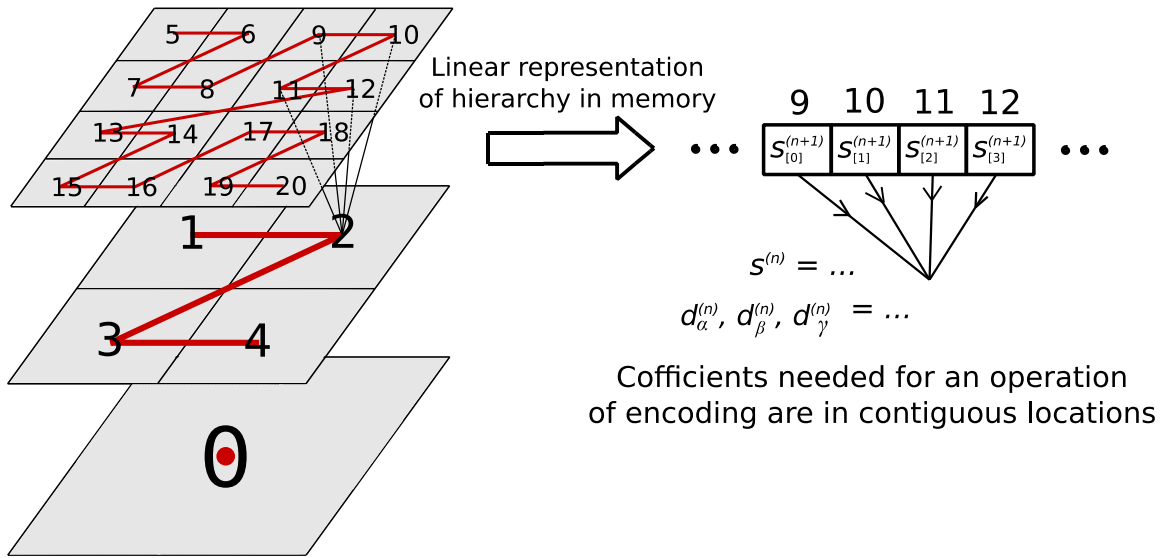

Figure 6: Z-order indexing of the hierarchy of grids so that each cell in the hierarchy has a unique Z-order index (left panel) and the corresponding locations of the children in memory (right panel).

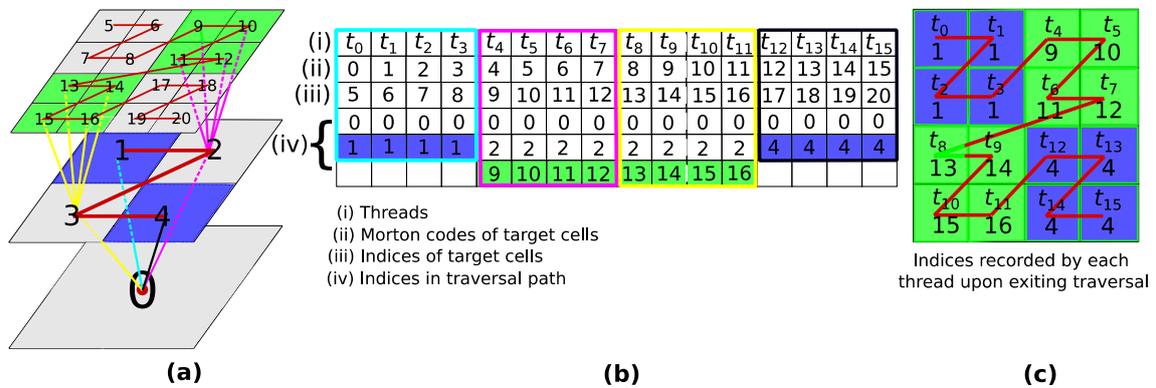

Figure 7: Parallel tree traversal (PTT). The left panel shows the tree of significant details after enforcing Z-order indexing, with different traversal paths indicated in yellow, cyan, magenta and grey. The middle panel shows the traversal paths of each thread during the PTT in terms of the Z-order indices of the cells they traverse. The right panel shows the Z-order indices recorded by each thread after PTT is complete.

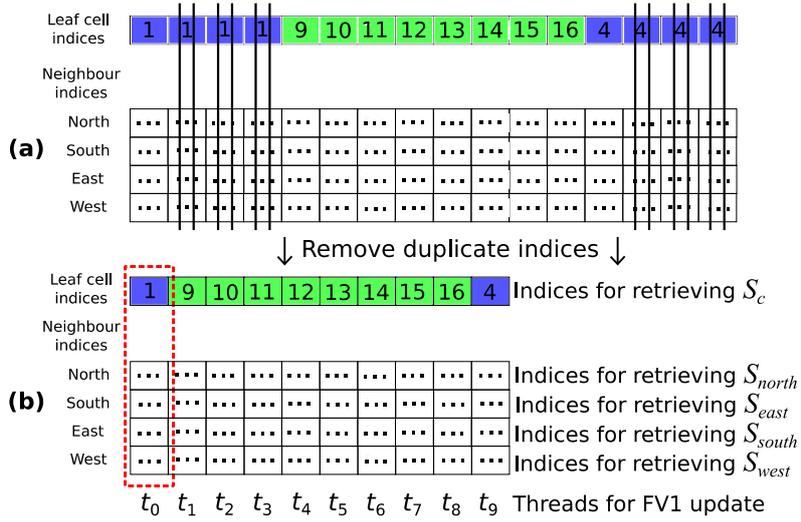

Figure 8: Parallel FV1 update. Top panel shows the Z-order indices of the leaf cells and their neighbours stored in memory after PTT and neighbour finding. Bottom panel shows the Z-order indices of the leaf cells and their neighbours without any duplicates, used to retrieve $S_c$, $S_{west}$, $S_{east}$, $S_{north}$ and $S_{south}$ to compute $\mathbf{L}_c$ and perform the FV1 update.

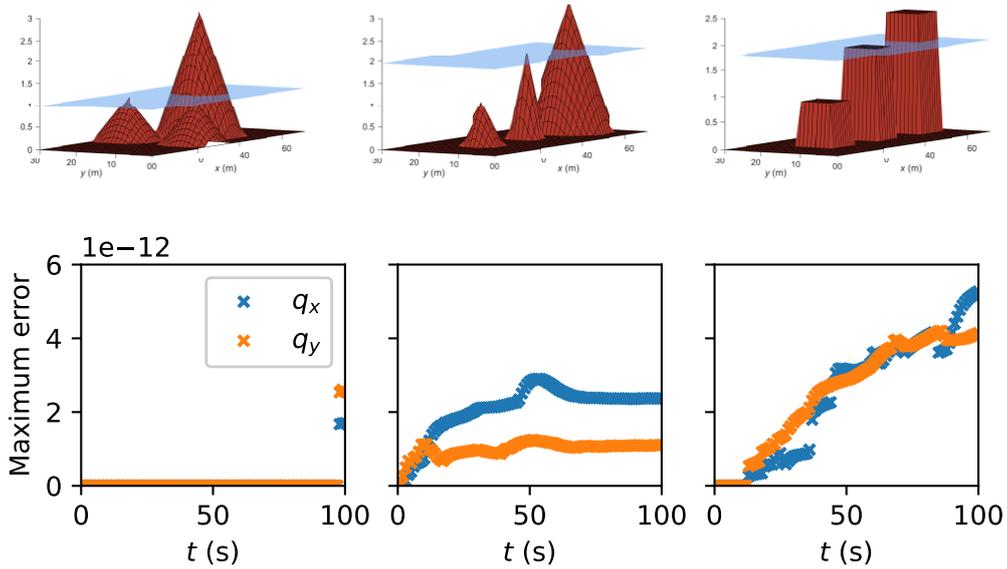

Figure 9: Verification of fidelity. Well-balanced property verification in the presence of wet-dry fronts with different levels of steepness in topography and different wetting conditions: smooth humps (left panels), steeper humps (middle panels) and rectangular humps (right panels). The top panels show the geometrical profiles of the humps in the domain area, and the lower panels include the time history of the maximum discharge errors where $q_x = hu$ and $q_y = hv$.

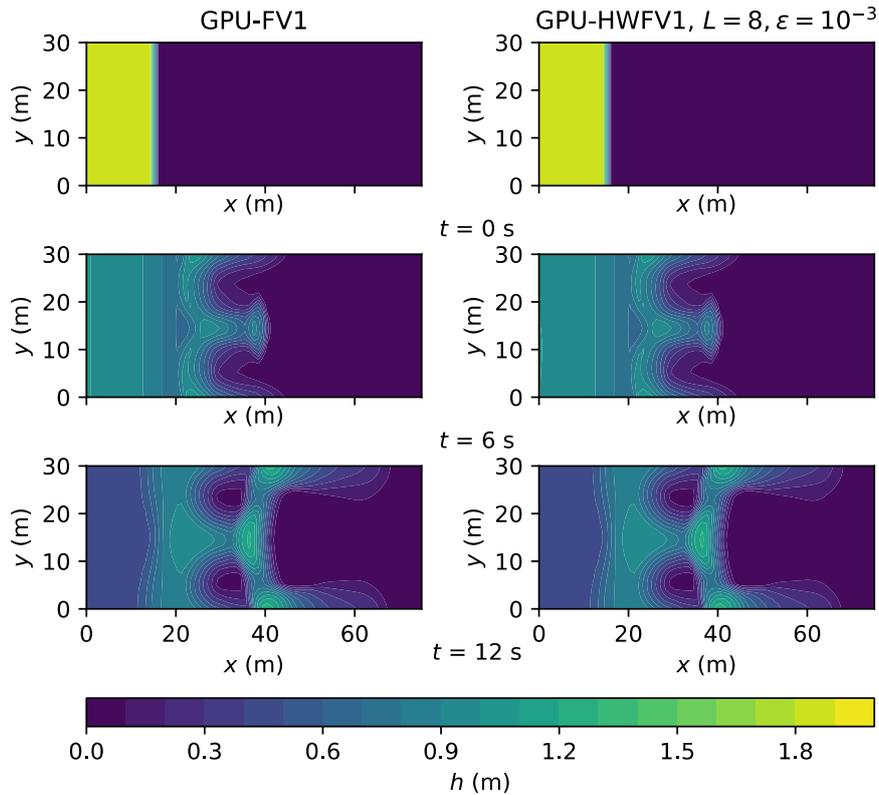

Figure 10: Verification of fidelity. Realistic dam-break flow with friction effects and moving wet-dry fronts. Flood depth profiles predicted by GPU-HWFV1 and GPU-FV1 on the left and right panels, respectively. At 0 s, the dam break wave emerges (top panels). At 6 s, the wave has submerged the small humps (middle panels). At 12 s, the wave starts to surround the large hump (bottom panels).

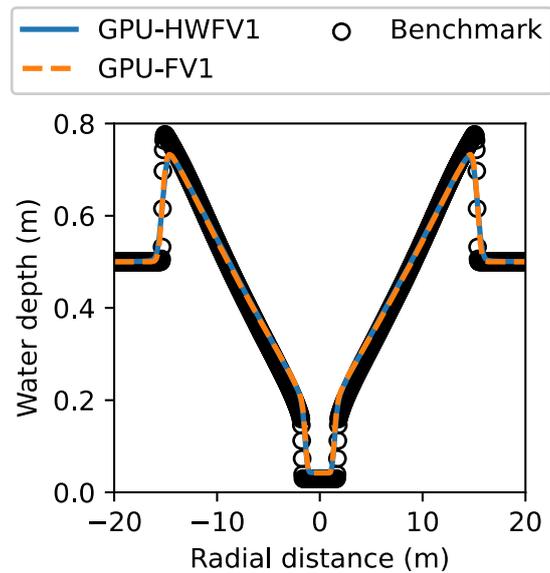

Figure 11: Circular 2D dam-break flow. Verification of GPU-HWFV1 using the same choice of $\varepsilon$ and $L$ as in Section 3.1 ($L = 8$ and $\varepsilon = 10^{-3}$): water depth centrelines at 3.5 s predicted by GPU-HWFV1, CPU-HWFV1, and GPU-FV1 compared to the benchmark solution.

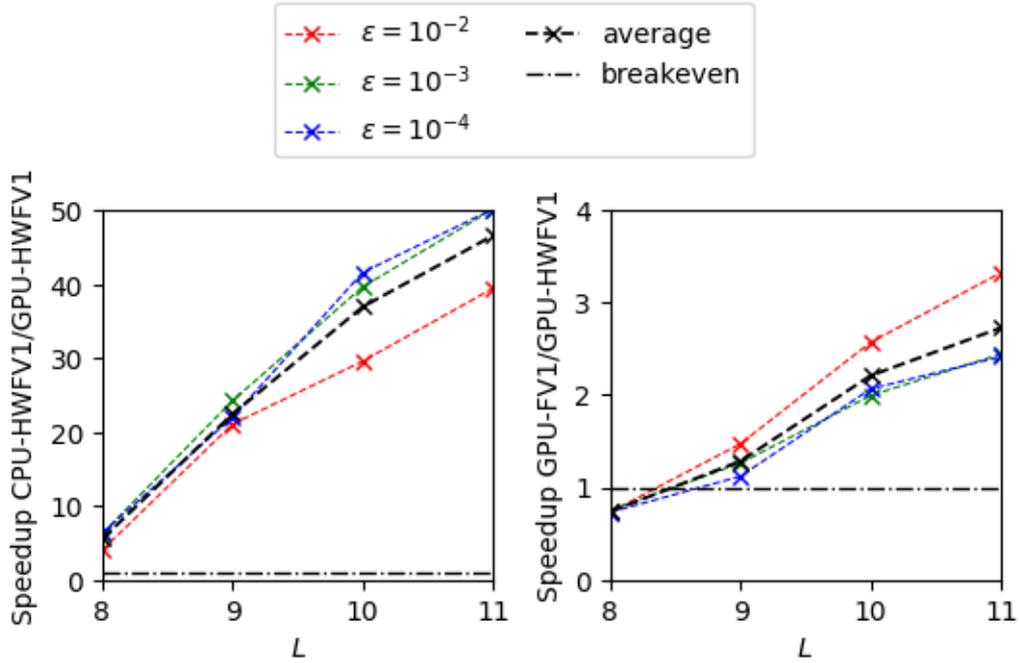

Figure 12: Circular 2D dam-break flow. Speed-up ratios to accomplish a 3.5 s simulation: GPU-HWFV1 over CPU-HWFV1 (left panel) and over GPU-FV1 (right panel).

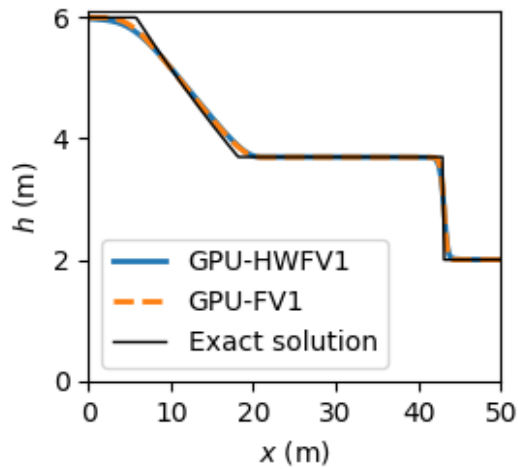

Figure 13: Pseudo-2D dam-break flow. Verification of GPU-HWFV1 using the same choice of $\varepsilon$ and $L$ as in Section 3.1 ($L = 8$ and $\varepsilon = 10^{-3}$): Water depths centerlines predicted by GPU-HWFV1, CPU-HWFV1 and GPU-FV1 at 2.5 s compared with the exact solution.

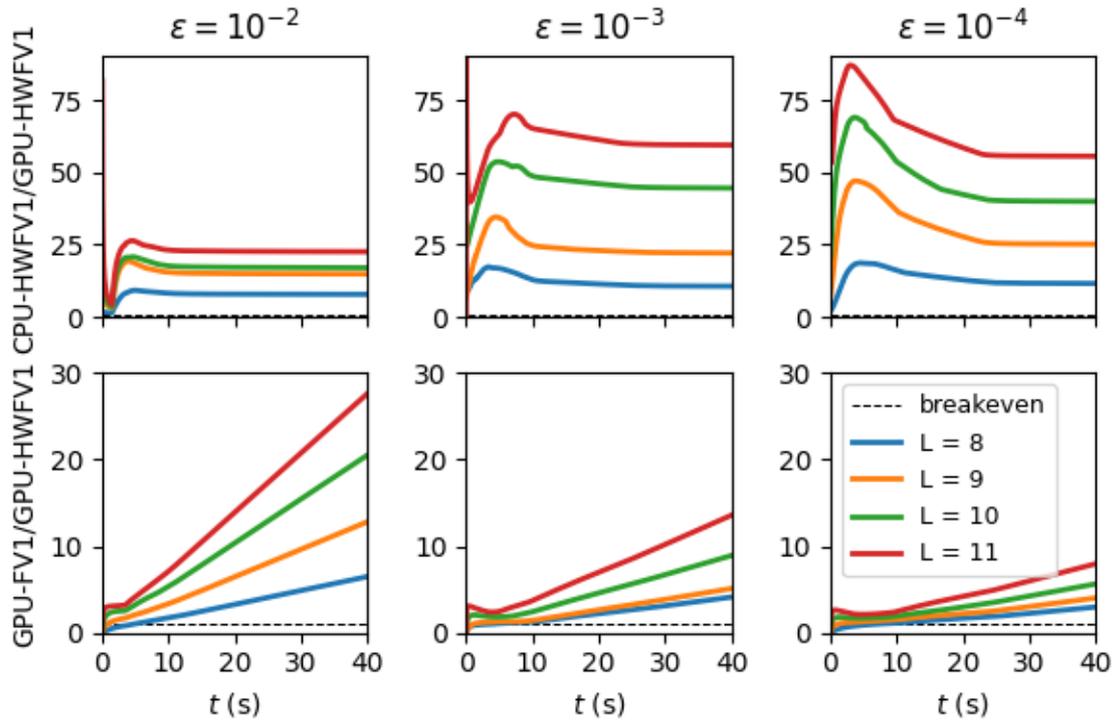

Figure 14: Pseudo-2D dam-break flow. Speed-up ratios of GPU-HWFV over CPU-HWFV1 (top panels), and over GPU-FV1 (bottom panels), for the three values of the threshold error $\varepsilon$ and considering different maximum refinement $L$.

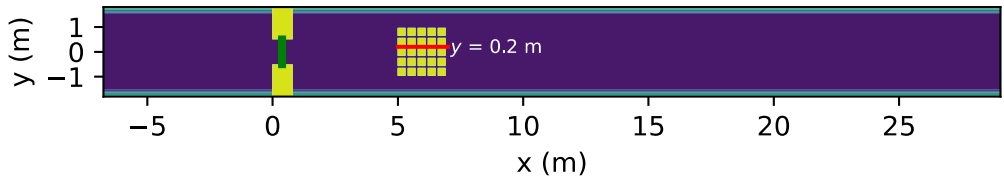

Figure 15: Dam-break wave interaction with an urban district. Top down view of the smooth channel with the gate indicated in green and topographic blocks coloured in yellow. Experimental depth and velocity data available along $y = 0.2$ m, indicated in red.

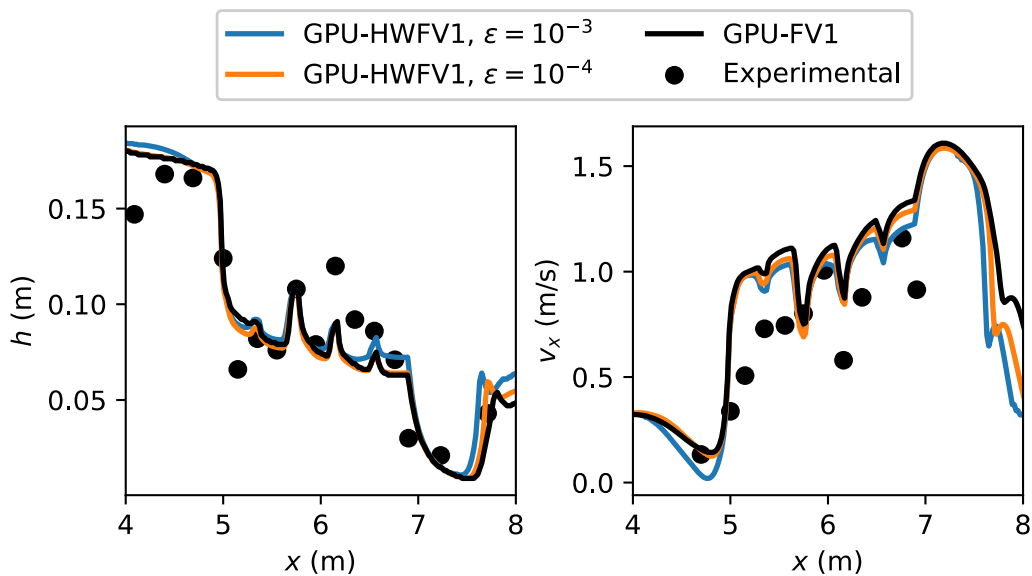

Figure 16: Dam-break wave interaction with an urban district. Depth (left panel) and velocity (right panel) profiles predicted along $y = 0.2$ m at 6 s by GPU-HWFV1 and GPU-FV1 compared to the experiments.

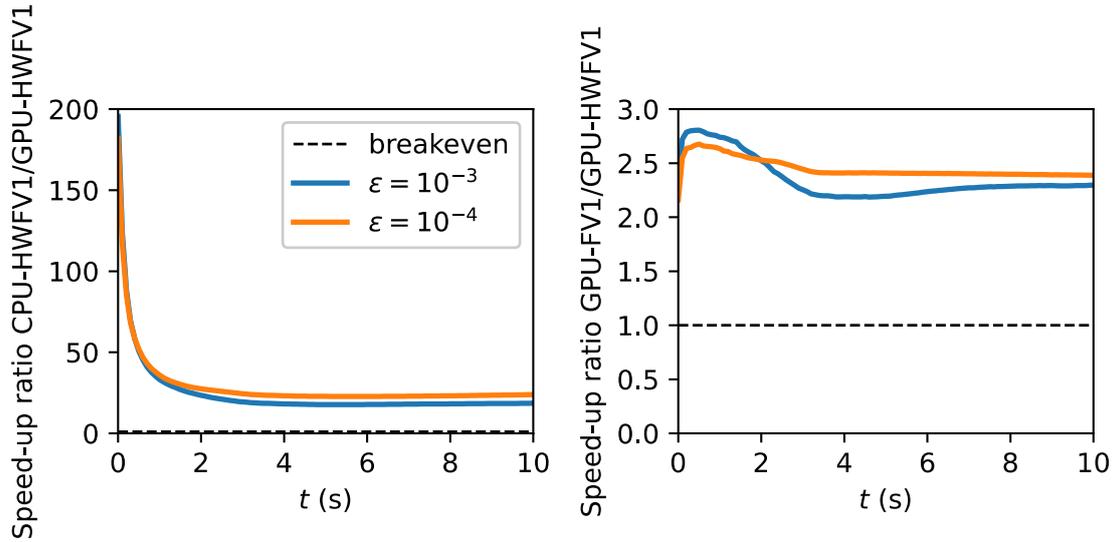

Figure 17: Dam-break wave interaction with an urban district. Speed-up ratios of GPU-HWFV1 over CPU-HWFV1 (left panel) and GPU-FV1 (right panel).

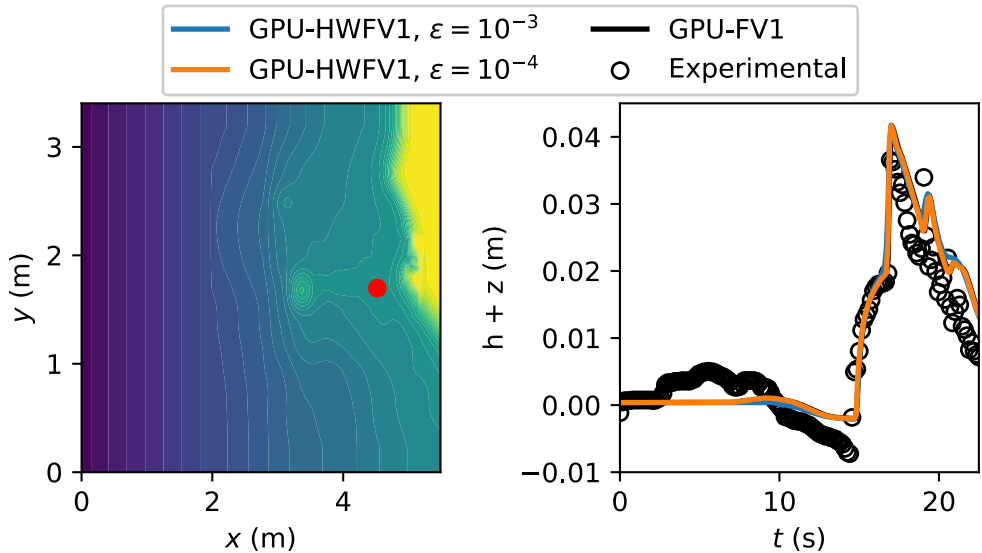

Figure 18: Tsunami wave propagation over a complex beach; (a) Topography contours over the domain area including the gauge point indicated in red. The tsunami-generated wave enters throughout the western boundary causing tsunami-generated flooding in the coastal area located in the eastern end (coloured in yellow); (b) Free-surface water elevation predicted by GPU-HWFV1 and GPU-FV1 compared to the experimental data.

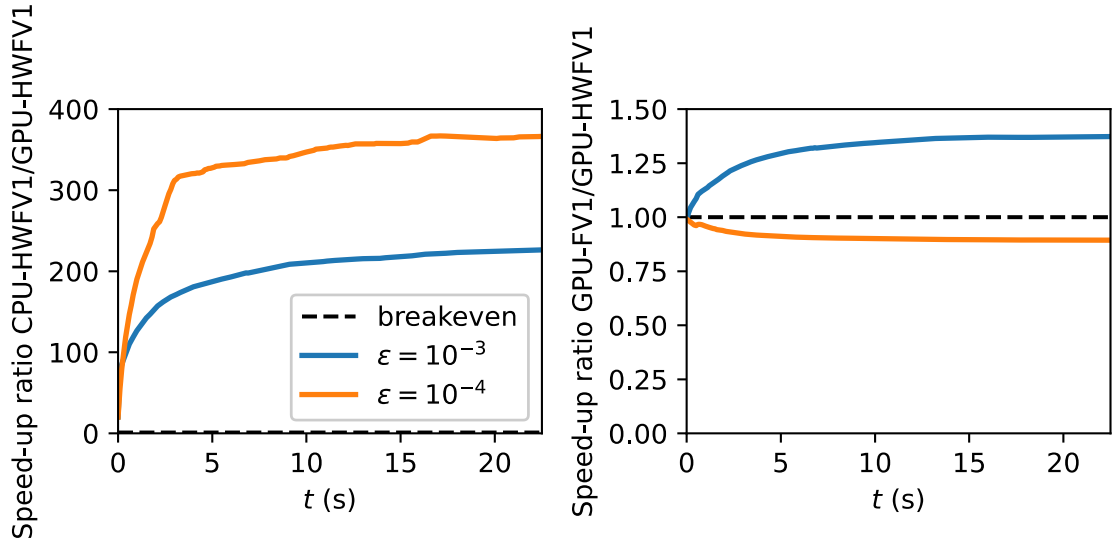

Figure 19: Tsunami wave propagation over a complex beach. Speed-up ratios of GPU-HWFV1 over CPU-HWFV1 (left panel) and over GPU-FV1 (right panel).